\documentclass[draftclsnofoot,onecolumn,12pt]{IEEEtran}

\usepackage[cmex10]{amsmath}    
\usepackage{stmaryrd}           
\usepackage{amsthm}             
\usepackage{amssymb}            
\usepackage{amsfonts}           

\usepackage{graphicx} 
\usepackage{cite} 
\usepackage{url}  
\usepackage{ifthen}  
\usepackage{multicol}   
\usepackage[utf8]{inputenc} 
\usepackage{enumerate} 
\usepackage{pdfpages} 
\usepackage{dblfloatfix} 
\usepackage{epstopdf}       
\usepackage{xcolor}
\usepackage{hyperref}

\theoremstyle{theorem}

\theoremstyle{proposition}
\newtheorem{proposition}{Proposition}
\theoremstyle{lemma}

\begin{document}

\title{Distributed Linear Precoding and User Selection in Coordinated Multicell Systems}
\author{Eduardo~Casta\~{n}eda,  Ad\~{a}o~Silva, Ramiro~Samano-Robles, and At\'{i}lio~Gameiro }


\markboth{Submitted to IEEE Transactions on Vehicular Technology}
{Submitted paper}

\maketitle

\begin{abstract}
In this manuscript we tackle the problem of semi-distributed user selection with distributed linear precoding for sum rate maximization in multiuser multicell systems.
A set of adjacent base stations (BS) form a cluster in order to perform coordinated transmission to cell-edge users, and coordination is carried out through a central processing unit (CU). However, the message exchange between BSs and the CU is limited to scheduling control signaling  and no user data or channel state information (CSI) exchange is allowed.
In the considered multicell coordinated approach, each BS has its own set of cell-edge users and transmits only to one intended user while interference to non-intended users at other BSs is suppressed by signal steering (precoding). We use two distributed linear precoding schemes, Distributed Zero Forcing (DZF) and Distributed Virtual Signal-to-Interference-plus-Noise Ratio (DVSINR).
Considering multiple users per cell and the backhaul limitations, the BSs rely on local CSI to solve the user selection problem. First we investigate how the signal-to-noise-ratio (SNR) regime and the number of antennas at the BSs impact the effective channel gain (the magnitude of the channels after precoding) and its relationship with multiuser diversity.
Considering that user selection must be based on the type of implemented precoding, we develop metrics of compatibility (estimations of the effective channel gains) that can be computed from local CSI at each BS and reported to the CU for scheduling decisions. Based on such metrics, we design user selection algorithms that can find a set of users that potentially maximizes the sum rate. Numerical results show the effectiveness of the proposed metrics and algorithms for different configurations of users and antennas at the base stations.
\end{abstract}


\section{Introduction}
\label{sec:introduction}

\IEEEPARstart{T}{he} performance of coordinated downlink transmission with linear precoding in multiple antenna multicell systems has been an active area of research over the last years. Recent works (e.g. \cite{Hossain2011,Marsch2011,Nguyen2014} and references therein) have shown that  cooperation and coordination between clustered base stations (BSs) improve rates, coverage, and efficiently suppress inter-cell interference (ICI) which specially benefits cell-edge users \cite{Zhang2009}.
Multicell coordination involves message exchange between neighboring cells and according to the level of coordination, multicell systems have been classified in three groups \cite{Park2012a,Bjornson2013,Yang2013a,Nguyen2014}: interference aware (IA), joint processing/transmission (JT), and coordinated beamforming (CBF).
In IA there is no information exchange among BSs, each transmitter serves its own set of users, and transmission parameters are adjusted in a selfish fashion by measuring ICI \cite{Nguyen2014}.
In contrast, in JT systems it is assumed that channel state information (CSI) and user data are globally available, full coordination is attainable though a central processing unit (CU), and each user receives data from a group of coordinated BSs (cluster).
The JT system can be interpreted as a broadcast channel \cite{Bjornson2013} with distributed antennas and several radio resource management (RRM) tasks (e.g., scheduling, power control, precoding design, data queue control, etc.) extended from the single-cell systems can be applied (e.g., \cite{Zhang2009,Bjornson2010,Marsch2011,Hossain2011,Huh2012,Hong2012,Niu2013,Nguyen2014}). However, such extensions must take into account backhaul rate limitations, CSI acquisition, joint transmission, and other system constraints \cite{Yang2013a}.

In CBF the BSs need only data of the users in their own cells and they do not require to know the precoders and traffic of other BSs. The shared information is related to scheduling control signaling and CSI in order to mitigate spatial ICI. The BSs design precoding vectors towards the scheduled users so that the gain is two-fold: increasing the signal strength at the receivers and suppressing interference in the adjacent cells \cite{Yang2013a}. Efficient RRM schemes can be implemented under CBF using local CSI \cite{Silva2011,Yu2013} which relaxes the wideband backhaul and synchronization requirements \cite{Hossain2011}.

Regardless the type of coordination between neighboring BSs, the inter-cluster interference problem arises if multiple clusters are taken into account, which can be dealt in two ways. The most straightforward way is to apply the principle of cellular planning with frequency reuse \cite[\S 5]{Hossain2011}. Using different radio resources in adjacent clusters (it can be dynamically allocated) mitigates or eliminates the inter-cluster interference.
A second approach to reuse radio resource among different clusters is by means of inter-cluster coordination, where adjacent clusters implement interference mitigation techniques for the users at the edges of the clusters (e.g. \cite{Zhang2009}).
For sake of simplicity and modeling tractability this work considers a single cluster network with $B$ BSs for the single carrier case.

\subsection{Related Works}
\label{sub_sec:related_works}

Depending on the system utility function that is optimized, there exist different strategies to achieve optimal power allocation and precoding design assuming that global CSI is known, and that the number of antennas at the transmitters can serve all competing users (cf. \cite{Hong2012,Nguyen2014} for an in-depth survey).
In the scenario where each BS serves only one user and CSI is not exchanged among BSs, the system model can be referred to as \textit{interference channel} \cite{Bjornson2013}. Recent works characterize its achievable rate region and jointly perform power allocation and precoding design (e.g., \cite{Larsson2008,Liu2011,Park2012a,Hong2012}) under the assumption that the intended user of each BS has been previously selected by some procedure.
However, for multiuser multicell scenarios each BS must select one user from its own pool of users before proceeding with precoding calculations.  In this scenario the sum rate maximization is a complex combinatorial problem because the number of users is larger than the number of available spatial resources (antennas) and global CSI may not be available. The global performance is highly sensitive to the set of scheduled users, since the signal strength of the intended user $k_{b}$ at BS $b$ relies on local CSI (including the local channels of non-intended users at other BSs). Additionally, the multiplexing gain and ICI suppression depend upon the number of antennas at the BSs \cite{Yang2013a}.

In the literature of single-cell MU-MISO systems with precoding based on Zero Forcing (ZF),  the sum rates maximization problem is commonly tackled by decoupling the user selection from the power allocation and precoding design.
The user selection is performed first based on the null space projection (NSP) (e.g., \cite{Tu2003,Yoo2006,Chan2007}) or an approximation of it (e.g., \cite{Fuchs2007,Castaneda2014}). The NSP provides an accurate measure of the \textit{effective channel gain} (the channel magnitude after precoding), so that the user channels selected based on such metric are spatially compatible or quasi-orthogonal. For ZF precoding, this means that the users selected using the NSP can provide a close-to-optimal solution to the sum rate maximization problem in multiuser scenarios.
Recent works on multicell systems have proposed extensions from single-cell user selection algorithms assuming that partial or global CSI is available at the CU (e.g., \cite{Zhang2009,Marsch2011,Huh2012,Bjornson2013,Niu2013}).
The extensions in \cite{Zhang2009,Marsch2011} are centralized algorithms that exploit the concept of NSP to improve sum rates relying on global CSI at the scheduler.
If global CSI is not available, distributed precoding and scheduling can still be implemented. For instance, LTE-Advanced standard \cite{Hossain2011,Lee2012a} considers distributed linear precoding such as signal-to-leakage-plus-noise ratio (SLNR) \cite{Sadek2007,Bjornson2013} and ZF whose computation requires to know only local CSI and the set of intended users.
One strategy for joint distributed precoding and scheduling is to limit the exchange of CSI such that the clustered BSs jointly select users in a sequential fashion, i.e., the first BS selects its user and broadcast its decision, then the second BS selects its user based on the decision made by the first one and so on \cite{Hossain2011}. Another approach has been introduced in \cite{Yu2013} where users selection, precoding design, and power allocation are treated as decoupled problems but their parameters are jointly updated at the CU. Results show that distributed RRM schemes with limited message exchange between BSs can improve system performance.

\subsection{Contributions}
\label{sub_sec:contributions}

In the system model considered in this work, a set of adjacent BSs form a cluster and they coordinate their transmission strategies through a CU in order to serve a set of cell-edge users and mitigate ICI.
The clustered BSs adopt the CBF transmission scheme where the data for an intended user is transmitted from one BS, whereas the impairments from the ICI are mitigated by coordinated precoding.
Two distributed linear precoding schemes will be used: Distributed Zero Forcing (DZF) and Distributed Virtual Signal-to-Interference-plus-Noise Ratio (DVSINR derived from SLNR).
It is assumed that each BS has its own set of intended users, no user data or CSI is exchanged between BSs, and the shared information between BSs and the CU is for scheduling control. In each scheduling instance the clustered BSs attempt to maximize the sum rate by selecting a set of users with particular characteristics.
Optimizing the performance in the described scenario is a challenging task since global CSI is not available and the backhaul connection with the CU only supports scheduling control information. Moreover, selecting the best set of users whose channel characteristics maximize the sum rate is a combinatorial problem whose complexity grows exponentially with the number of BSs and users per cell \cite{Yu2013}.

To solve the user selection problem, and taking into account that the BSs implement either DZF or DVSINR, the key results of this work are summarized as follows.

$\bullet$ Initially, we discuss how the instantaneous and average effective channel gains of DZF and DVSINR depend on the signal-to-noise-ratio (SNR) regime, the number of antennas at the BSs, and multiuser diversity. This insight of the precoder schemes is used to establish in which way local CSI must be processed at each BS. We design precoder-based metrics of user compatibility, i.e., depending on the type of precoding we propose a mapping from the local CSI to a real number. The proposed metrics are estimations of the achievable effective channel gains and operate in different system configurations based on the number of transmit antennas and BSs.

$\bullet$ The scheduling process must be perform at the CU using the metrics reported by the BSs. We accomplish this goal by developing an algorithm for user selection that properly combines the reported metrics. Once that a set of users has been selected, the decision is informed to the BSs and they compute either DZF or DVSINR based on the local CSI of the selected users.

$\bullet$ We propose a pre-selection methodology in order to reduce the number of competing users per BS. The method is a ranking-based per-antenna selection that preserves multiuser diversity in CBF systems and reduces the amount of information exchanged between the BSs and the CU. Numerical results show that our proposed metrics and algorithms for user selection can achieve a large portion of the optimal sum rate (the benchmark is a fully centralized system) by exploiting local CSI with limited message exchange between BSs and the CU.

The remainder of the paper is organized as follows. The system model and the problem formulation are presented in Section~\ref{sec:system_model}.
In Section~\ref{sec:distributed_precoders} we present the DZF and DVSINR precoding schemes, their properties, expressions for their effective channel gains, and their relation with user selection.
In Section~\ref{sec:metrics_spatial_compatibility} we define the metrics that estimate the effective channel gains and Section~\ref{sec:multicell_user_selection} presents the semi-distributed user selection algorithm whose solution set solves the sum rate maximization problem. Numerical results are provided in Section~\ref{sec:numerical_results} and conclusions are drawn in Section~\ref{sec:conslusions}.

Notation: matrices and vectors are set in upper and lower boldface respectively.
$\left<\mathbf{a},\mathbf{b}\right>$ is the inner product between vectors $\mathbf{a}$ and $\mathbf{b}$. $(\cdot)^{T}$, $(\cdot)^{H}$, $|\cdot|$, $\|\cdot\|$ denote the transpose, hermitian transpose, absolute value, and vector norm respectively. Calligraphic letters, e.g. $\mathcal{G}$, denote sets and $|\mathcal{G}|$ denotes cardinality. $Tr(\cdot)$, and $\det(\cdot)$ represent the trace and determinant operators. $\mathbb{E}[ \cdot ]$ represents the expectation operation. $Sp(\mathbf{A})$ and $Sp(\mathbf{A})^{\perp}$ denote the subspace and orthogonal subspace spanned by the columns of matrix $\mathbf{A}$.
$\lambda_{i}(\mathbf{A})$ is the $i$th eigenvalue of the operated matrix, $\lambda_{\max}(\mathbf{A})$, $\lambda_{\min}(\mathbf{A})$, $rank(\mathbf{A})$, $null(\mathbf{A})$ are the maximum and minimum eigenvalues, rank and null space of matrix $\mathbf{A}$ respectively. $\mathbf{eig}(\mathbf{A}) = [\lambda_{i},\ldots,\lambda_{n}]$ is the vector that contains all $n$ eigenvalues of matrix $\mathbf{A}$. Let $\mathbf{x}$ be a vector, then $[\mathbf{x}]_{i} = x_{i}$ is the $i$th element. $\mathbf{I}_{n}$ is the identity matrix of size $n$. $\mathbb{R}_{+}$ is the set of nonnegative real numbers.
For a given a vector $\mathbf{x} \in \mathbb{R}_{+}^{n}$, the Jain's index of fairness is defined as follows \cite{Jain1984}:
\begin{equation}\label{eq:jains_index}
    J(\mathbf{x}) \triangleq \frac{\left( \sum_{i=1}^{n} x_i \right)^2}{ n\sum_{i=1}^{n} x_i^2 },
\end{equation}
where $\{J(\cdot) \in \mathbb{R}_{+} | \frac{1}{n} \leq J(\cdot) \leq 1 \}$.

\section{Problem Formulation}
\label{sec:system_model}

We consider a multiuser multicell clustered network where a group of $B$ adjacent BSs form a cluster. Each BS has $N_{t}$ antennas, all users in the network are equipped with single antenna devices, and define $\epsilon \triangleq \max\{N_{t} - (B-1), 0\}$.
The BSs only exchange messages of scheduling control through a CU and precoding design is performed at each BS using local CSI.
The joint user selection and precoding design are performed for cell-edge users located in the cell-edge area defined by $B$ BSs. The users are deployed within a circular area that spans a radius $r_{coop}$ (a fraction of the cell radius $r$).
The $b$th BS has one index set of edge users $\mathcal{S}_{b}$ and it only transmits data to one user in this set.
Consider that $\mathcal{S}_{b} \cap \mathcal{S}_{j} = \emptyset$, $\forall j \neq b$ and the transmitted signal from BS $b$ to user $k_{b} \in \mathcal{S}_{b}$ is: $\mathbf{x}_{b} = \sqrt{P_{b}} \mathbf{w}_{b} s_{b}$. $P_{b}$ is the transmitted power, $\mathbf{w}_{b} \in \mathbb{C}^{N_t \times 1}$ is the unit norm precoder and $s_{b}$ is the transmitted data symbol with $\mathbb{E}[|s_{b}|^2]=1$,  $\mathbb{E}[\|\mathbf{x}_{b}\|^2]=P_{b}$, and $P_{b}\leq P$ where $P$ is the maximum available power.
The received signal of the intended user $k_{b}$ of BS $b$ is given by:

\begin{equation} \label{eq:received_signal_no_data_sharing}
    y_{bk_{b}} = \sqrt{P_{b}}\mathbf{h}_{bk_{b}}^{H} \mathbf{w}_{b} s_{b}
    + \sum_{j=1, j \neq b}^{B} \sqrt{P_{j}}\mathbf{h}_{jk_{b}}^{H}\mathbf{w}_{j}s_{j} + n_{k_{b}},
\end{equation}
where $\mathbf{h}_{bk_{b}}\sim \mathcal{CN}(0,\varrho_{bk_{b}}^{2}\mathbf{I})$ of size $N_t \times 1$ is a flat Rayleigh fading propagation channel between user $k_{b}$ and BS $b$ and $\varrho_{bk_{b}}^{2}$ is the long-term channel power gain.  The term $n_{k_{b}}\sim \mathcal{CN}(0,\sigma_{n}^{2})$ is the noise.
The receivers treat co-terminal interference as noise and the instantaneous signal-to-interference-plus-noise ratio (SINR) of user $k_{b} \in \mathcal{S}_{b}$ is defined as:
\begin{equation}\label{eq:instantaneous_sinr}
    SINR_{bk_{b}} = \frac{ {P_{b}}|\mathbf{h}_{bk_{b}}^{H}\mathbf{w}_{b}|^{2} } { \sum_{j=1, j\neq b}^{B} {P_{j}} |\mathbf{h}_{jk_{b}}^{H}\mathbf{w}_{j}|^{2} + \sigma_{n}^{2} }.
\end{equation}

In a cluster with $B$ BSs, there exists $L = \prod_{b=1}^{B} |\mathcal{S}_{b}|$ user permutations of $B$ users that can be chosen for simultaneous transmission.
Each user in $\mathcal{S} = \bigcup_{b=1}^{B} \mathcal{S}_{b}$ has a unique index and all BSs know which indices belong to each BS.
Let $\mathcal{G}_{l}$ $\forall l \in \{1,\ldots,L\}$ be a set of $B$ users where each user is served by one BS and the users indices in the set $l$ are the same for all BSs.
The set $\mathcal{G}_{l}$ has an associated channel matrix at the $b$th BS which is given by $\mathbf{H}_{b}^{(l)} \triangleq \{ \mathbf{h}_{bk_{i}}: k_{i} \in \mathcal{G}_{l} \}$, i.e., all the local channels of the users grouped in $\mathcal{G}_{l}$.
We need to solve the sum rate maximization problem in the multiuser multicell scenario defined as:

\begin{equation} \label{eq:sum_rate_max_problem}
\begin{IEEEeqnarraybox}[
\IEEEeqnarraystrutmode
\IEEEeqnarraystrutsizeadd{2pt}
{2pt}
][c]{rCl}
\underset{l \in \{1,\ldots,L\}}{\text{maximize}} & \hspace{8pt} & \sum_{b=1}^{B} \log_{2}\left(1+SINR_{bk_{b}} \right)
\\
\text{subject to} &  & \|\mathbf{w}_{b}^{(type)}(\mathbf{H}_{b}^{(l)})\|^{2} = 1,  \ \ \forall b \in \{ 1, \ldots, B \}
\end{IEEEeqnarraybox}
\end{equation}
where $SINR_{bk_{b}}$ is defined in (\ref{eq:instantaneous_sinr}) for the user $k_{b} \in \{ \mathcal{S}_{b} \cap \mathcal{G}_{l} \}$. The precoding vectors $\mathbf{w}_{b}^{(type)}(\mathbf{H}_{b}^{(l)})$ $\forall b$ are functions of $\mathbf{H}_{b}^{(l)}$ at each BS for the given set $l$, and  $type \in \{DZF,DVSINR\}$ is the implemented precoding technique which will be defined in the next section.
Our objective is to find the set $l$ that solves problem (\ref{eq:sum_rate_max_problem}) which can be attained  by taking advantage of the properties of $\mathbf{w}_{b}^{(type)}$. Such properties are used to exploit the local CSI in order to evaluate the effective channel gains, i.e., $|\mathbf{h}_{bk_{b}}^{H}\mathbf{w}_{b}|^{2}$ which are tightly related with the achievable rates.

\section{Distributed Linear Precoding}
\label{sec:distributed_precoders}

In this section we investigate two precoding techniques DZF and DVSINR. We need to define underlying characteristics of the precoders, their dependence on the SNR regime and $N_t$, and quantify how those characteristics affect the instantaneous and average effective channel gains.

\subsection{Distributed Zero Forcing (DZF)}
\label{sub_sec:dzf_precoder}

Zero-forcing is a precoding strategy that removes the inter-user interference and is defined always that $N_t\geq B$.
The conditions to achieve near Pareto-optimal rates with distributed ZF for the two-BS scenarios were presented in \cite{Larsson2008} and for $B$ BSs generalized expressions to compute $\mathbf{w}_{b}$ are provided in \cite{Bjornson2013,Silva2011}.
Let $\tilde{\mathbf{H}}_{bk_{b}}$ be the aggregate interference matrix of user $k_{b}$ given by:

\begin{equation}\label{eq:aggregated_interference_matrix}
    \tilde{\mathbf{H}}_{bk_{b}} = [ \mathbf{h}_{bk_{1}}, \ldots, \mathbf{h}_{bk_{b-1}}, \mathbf{h}_{bk_{b+1}}, \ldots, \mathbf{h}_{bk_{B}} ],
\end{equation}
and each term $\mathbf{h}_{bk_{i}}$ $\forall i\neq b$ corresponds to the channel between BS $b$ and the non-intended user $k_{i} \in \mathcal{S}_{i}$.
The matrix $\tilde{\mathbf{V}}_{\tilde{\mathbf{H}}_{bk_{b}}} = null(\tilde{\mathbf{H}}_{bk_{b}})$ contains $\epsilon$ column vectors\footnote{Using SVD $\tilde{\mathbf{H}}_{bk_{b}} = [ \bar{\mathbf{W}}_{\tilde{\mathbf{H}}_{bk_{b}}} \tilde{\mathbf{W}}_{\tilde{\mathbf{H}}_{bk_{b}}} ] \boldsymbol\Sigma_{\tilde{\mathbf{H}}_{bk_{b}}} \mathbf{O}_{\tilde{\mathbf{H}}_{bk_{b}}}^{H}$ where $\tilde{\mathbf{W}}_{\tilde{\mathbf{H}}_{bk_{b}}}$ contains $\epsilon$ orthonormal vectors that are the basis of the null space of $\tilde{\mathbf{H}}_{bk_{b}}$ and $\tilde{\mathbf{V}}_{\tilde{\mathbf{H}}_{bk_{b}}} = null(\tilde{\mathbf{H}}_{bk_{b}}) = \tilde{\mathbf{W}}_{\tilde{\mathbf{H}}_{bk_{b}}}$.} that are candidates to form $\mathbf{w}_{b}$ since they will produce zero interference to the other users in $\tilde{\mathbf{H}}_{bk_{b}}$.
If $\epsilon>1$ the elements of $\tilde{\mathbf{V}}_{\tilde{\mathbf{H}}_{bk_{b}}}$ can be linearly combined to form the precoding vector as follows \cite{Silva2011}:

\begin{equation}\label{eq:define_dzf_precoder}
    \mathbf{w}_{b}^{(DZF)} = \tilde{\mathbf{V}}_{\tilde{\mathbf{H}}_{bk_{b}}} \frac{ ( \mathbf{h}_{bk_{b}}^{H} \tilde{\mathbf{V}}_{\tilde{\mathbf{H}}_{bk_{b}}} )^{H} } { \| \mathbf{h}_{bk_{b}}^{H} \tilde{\mathbf{V}}_{\tilde{\mathbf{H}}_{bk_{b}}} \| },
\end{equation}
and the received signal at user $k_{b}$ has its phase aligned.

\begin{proposition}[]\label{thm:1}
    The expected value of the effective channel gain of the intended user $k_{b} \in \mathcal{S}_{b}$ served by BS $b$ using DZF precoding with $\mathbf{w}_{b}$ defined in (\ref{eq:define_dzf_precoder}) under constraint $N_t\geq B$ is defined as follows:
    \begin{equation}
        \mathbb{E}\left[ |\mathbf{h}_{bk_{b}}^{H}\mathbf{w}_{b} |^2 \right]
        = \frac{\epsilon}{N_t}  \mathbb{E} \left[  \|\mathbf{h}_{bk_{b}}\|^2  \right] \label{eq:theorem_upper_bound_average_rate_dzf}
    \end{equation}
\end{proposition}

\begin{IEEEproof}
See Appendix~\ref{appendix:1}
\end{IEEEproof}

\subsection{Distributed Virtual SINR (DVSINR)}
\label{sub_sec:dvsinr_precoder}

The ideal precoder technique would be able to balance between signal power maximization and interference power minimization and a heuristic way to find such balance is reached by maximizing the SLNR \cite{Bjornson2013}. In \cite{Bjornson2010} the authors show that it is possible to achieve Pareto-optimal rates in multicell transmission when the precoding vectors are given by:

\begin{equation}\label{eq:define_general_dvsinr_precoder}
    \mathbf{w}_{b}^{\star} = \arg \underset{\|\mathbf{w}\|^{2}=1}{\max} \frac{ \upsilon_{bk_{b}}|\mathbf{h}_{bk_{b}}^{H} \mathbf{w}|^2 }{  \sum_{j=1, j\neq b}^{B} \upsilon_{bk_{j}}|\mathbf{h}_{bk_{j}}^{H} \mathbf{w}|^2 + \frac{\sigma_{n}^{2} }{ P_{b} }},
\end{equation}
where $\upsilon_{bk_{b}} \in (0,1)$ which is a heuristic extension of the SLNR precoding \cite{Sadek2007,Bjornson2013}. Then $\mathbf{w}_{b}^{\star}$ $\forall b$ are linear combinations of the maximal ratio transmission and ZF precoders and the coefficients $\upsilon_{bk_{b}}$ that optimally maximize the sum rate can be only computed with global CSI. If maximum ICI is accounted\footnote{The authors in \cite{Park2012a} showed that the coefficients $\upsilon_{bk}$ can define user weights that may represent, for instance, user priority.} $\upsilon_{bk_{b}}=1$ $\forall b$ the precoders that solve the virtual SINR maximization problem (\ref{eq:define_general_dvsinr_precoder}) are given by \cite{Bjornson2010}:

\begin{equation}\label{eq:define_dvsinr_precoder}
    \mathbf{w}_{b}^{(DVSINR)} = \frac{ \mathbf{D}_{bk_{b}}\mathbf{h}_{bk_{b}} }{ \| \mathbf{D}_{bk_{b}}\mathbf{h}_{bk_{b}} \| },
\end{equation}
where $\mathbf{D}_{bk_{b}} = \mathbf{C}_{bk_{b}}^{-1}$, $\mathbf{C}_{bk_{b}} \triangleq  \rho_{b}^{-1} \mathbf{I}_{N_{t}} + \tilde{\mathbf{H}}_{bk_{b}}\tilde{\mathbf{H}}_{bk_{b}}^{H}$ is a $N_{t} \times N_{t}$ positive-definite Hermitian matrix and $\rho_{b} = \frac{ P_{b} }{ \sigma_{n}^{2} }$.
The following result describes the relation between the eigenvalues of $\mathbf{D}_{bk_{b}}$ and the expected value of the effective channel gain.

\begin{proposition}[]\label{thm:2}
     The effective channel gain of the user $k_{b} \in \mathcal{S}_{b}$ served by the $b$th BS under DVSINR precoding constrained by $N_t\geq B$ can be bounded as follows:

    \begin{equation}\label{eq:theorem_upper_bound_average_rate_dvsinr}
        \mathbb{E} \left[ |\mathbf{h}_{bk_{b}}^{H}\mathbf{w}_{b} |^2 \right] \approx \mathbb{E} \left[ \|\mathbf{h}_{bk_{b}}\|^2  J(\mathbf{eig}(\mathbf{D}_{bk_{b}}))  \right]
    \end{equation}
    where $\mathbf{w}_{b}$ is defined in (\ref{eq:define_dvsinr_precoder}) and $J(\cdot)$ is the Jain's fairness index.
\end{proposition}
\begin{IEEEproof}
See Appendix~\ref{appendix:2}
\end{IEEEproof}

\begin{proposition}[]\label{thm:3}
    For DVSINR, given the matrix $\tilde{\mathbf{H}}_{bk_{b}} \in \mathbb{C}^{N_t \times (B-1)}$ and its corresponding $\mathbf{D}_{bk_{b}} \in \mathbb{C}^{N_t \times N_t}$ under constraint $N_t\geq B$ it holds that
    \begin{equation} \label{eq:theorem_dvsinr_high_snr_limit}
        \lim_{\rho_{b} \rightarrow \infty} \left( J(\mathbf{eig}(\mathbf{D}_{bk_{b}})) - \frac{\epsilon}{N_t}  \right) = 0,
    \end{equation}
    which implies that $\exists \rho_{0}$ and $\forall \rho_{b} \geq \rho_{0}$ the expected value of the effective channel gain is upper bounded as follows:
    \begin{equation} \label{eq:theorem_dvsinr_high_snr_effective_channel_gain}
        \mathbb{E} \left[  |\mathbf{h}_{bk_{b}}^{H}\mathbf{w}_{b} |^2 \right]
        \leq \frac{\epsilon}{N_t} \mathbb{E} \left[ \|\mathbf{h}_{bk_{b}}\|^2  \right]
    \end{equation}
\end{proposition}
\begin{IEEEproof}
See Appendix~\ref{appendix:3}
\end{IEEEproof}

The ICI for DVSINR is nonzero and for the high SNR regime the interference components in the denominator of (\ref{eq:instantaneous_sinr}) are usually neglected \cite{Bjornson2010,Park2012a}. The following result provides an approximation of the power that is leaked from clustered BSs using DVSINR precoding.

\begin{proposition}[]\label{prop:noise_interference_ratio_for_dvsinr}
    For DVSINR and $N_t \geq B$, the magnitude of the interference or leakage from the $j$th BS over the channel $\mathbf{h}_{jk_{b}} \in \tilde{\mathbf{H}}_{jk_{j}}$ $\forall j \neq b$ in the denominator of (\ref{eq:instantaneous_sinr}) for the user $k_{b} \in \mathcal{S}_{b}$ served by the $b$th BS can be aproximated as follows:
    \begin{equation} \label{eq:interference_component_dvsinr_approximation}
        \mathbb{E}\left[  |\mathbf{h}_{jk_{b}}^{H}\mathbf{w}_{j} |^2 \right] \approx \mathbb{E}\left[  \frac{ \|\mathbf{h}_{jk_{b}}\|^{2} }{ \epsilon  (\rho_{j}\lambda_{\min}(\tilde{\mathbf{H}}_{jk_{j}}^{H}\tilde{\mathbf{H}}_{jk_{j}}) + 1)^{2} } \right],
    \end{equation}
    where $\mathbf{w}_{j}$ is a function of the matrix $\tilde{\mathbf{H}}_{jk_{j}}$ associated to the user $k_{j}$ served by the $j$th BS.
\end{proposition}
\begin{IEEEproof}
See Appendix~\ref{appendix:3}
\end{IEEEproof}

\subsection{Distributed Linear Precoding and User Selection}
\label{sub_sec:precoding_and_user_selection_discussion}

Consider that $k_{b} \in \{ \mathcal{S}_{b} \cap \mathcal{G}_{l} \}$ and let $\tilde{\mathbf{H}}_{bk_{b}}(\mathcal{G}_{l}) \in \mathbb{C}^{N_{t} \times (B-1)}$ be the aggregate interference matrix of $k_{b}$ which contains all channels of $\mathbf{H}_{b}^{(l)}$ except $\mathbf{h}_{bk_{b}}$.
\subsubsection{DZF}
This scheme is defined if $N_t \geq B$ and achieves zero inter-user interference, i.e., $|\mathbf{h}_{bk_{j}}^{H}\mathbf{w}_{b}^{(DZF)} |^2=0$,  $\forall k_{j} \in \mathcal{G}_{l} \setminus \{k_{b}\}$.
From Proposition~\ref{thm:1} observe that the average value of $|\mathbf{h}_{bk_{b}}^{H}\mathbf{w}_{b}^{(DZF)}|^2$ depends on $N_t$ and $\epsilon$. As $\epsilon$ grows the effective channel gain is enhanced. However, results in Appendix~\ref{appendix:1} show that the instantaneous effective channel gain is a function of the angle between $\mathbf{h}_{bk_{b}}$ and the basis of the null subspace of $\tilde{\mathbf{H}}_{bk_{b}}(\mathcal{G}_{l})$.
The system performance is optimized, regardless the SNR regime, if the intended direct channel $\mathbf{h}_{bk_{b}}$ at BS $b$ maximize $|\mathbf{h}_{bk_{b}}^{H}\mathbf{w}_{b}^{(DZF)} |^2$. This means that channel magnitude and spatial compatibility (quasi-orthogonality w.r.t. $Sp(\tilde{\mathbf{H}}_{bk_{b}}(\mathcal{G}_{l}))$) must be optimized jointly.
Notice that $\forall k_{b} \in \mathcal{S}_{b}$ there exists $\prod_{j=1, j\neq b}^{B}|\mathcal{S}_{j}|$ possible precoders and the set $\mathcal{G}_{l}$ that maximizes  $|\mathbf{h}_{bk_{b}}^{H}\mathbf{w}_{b}^{(DZF)} |^2$ at BS $b$ is, in general, not the best set at other BSs.

\subsubsection{DVSINR}
This scheme does not impose a constraint on $N_t$ but its capacity to combat inter-user interference depends on it. For a given user set $\mathcal{G}_{l}$, Proposition~\ref{thm:2} shows that in the low and medium SNR regimes the expected value of $|\mathbf{h}_{bk_{b}}^{H}\mathbf{w}_{b}^{(DVSINR)}|^2$ depends on the magnitude of $\mathbf{h}_{bk_{b}}$ and the characteristics of $\tilde{\mathbf{H}}_{bk_{b}}(\mathcal{G}_{l})$. In particular, the magnitude of each i.i.d. vector in $\tilde{\mathbf{H}}_{bk_{b}}(\mathcal{G}_{l})$ and its singular values which directly modifies $J(\mathbf{eig}(\mathbf{D}_{bk_{b}}))$. In the high SNR, Proposition~\ref{thm:3} indicates that the expected value of $|\mathbf{h}_{bk_{b}}^{H}\mathbf{w}_{b}^{(DVSINR)}|^2$ is limited by $\epsilon$ and $N_t$ similar to DZF. Since the impact of $\mathbf{D}_{bk_{b}}$ in the effective channel gain is dominated by $\epsilon$ eigenvalues associated to the orthogonal subspace to $\mathcal{V}_{bk_{b}} =Sp(\tilde{\mathbf{H}}_{bk_{b}}(\mathcal{G}_{l}))$, the selected user at each BS should meet the same conditions previously described for DZF.
At the low SNR the eigenvalues of $\mathbf{D}_{bk_{b}}$ have similar magnitudes\footnote{From the definition in (\ref{eq:d_matrix_eigenvalues}) observe that for the low SNR regime $\rho^{-1} > [\boldsymbol\Sigma_{\hat{\mathbf{H}}}]_{ii}$, i.e., the eigenvalues of $\hat{\mathbf{H}}$ are negligible.} and the BS can select its user selfishly based on the channel magnitudes regardless the characteristics of $\mathcal{V}_{bk_{b}}$. At medium SNR the user selection is more complicated since the instantaneous effective channel gain is modified by the weighted basis of $\mathcal{V}_{bk_{b}}$ where the weights are functions of $\rho_{b}$ and $\epsilon$, cf. (\ref{eq:beta_of_h}) in Appendix~\ref{appendix:2}.
Notice that because $|\mathbf{h}_{bk_{j}}^{H}\mathbf{w}_{b}^{(DVSINR)} |^2\neq 0$,  $\forall k_{j} \in \mathcal{G}_{l} \setminus \{k_{b}\}$ the achievable SINR (\ref{eq:instantaneous_sinr}) strongly depends on $N_t$ and $B$.
If $\epsilon >0$ (power limited scenario) the amount of leaked power from BS $j$ to the user $k_{b}$ served by BS $b$ is scaled by a factor $\epsilon^{-1}$ as shown in (\ref{eq:interference_component_dvsinr_approximation}). When $\rho_{j} \rightarrow\infty$ the leakage is also scaled by a factor of $\rho_{j}^{-2}$ according to Proposition~\ref{prop:noise_interference_ratio_for_dvsinr} and inter-user interference vanishes. The expression (\ref{eq:interference_component_dvsinr_approximation}) reveals that for a fixed $\rho_{j}$ the leakage is minimized if $\lambda_{\min}(\tilde{\mathbf{H}}_{jk_{j}}^{H}\tilde{\mathbf{H}}_{jk_{j}})$ is maximized, which occurs if the i.i.d. vectors in $\tilde{\mathbf{H}}_{jk_{j}}$ are quasi-orthogonal.
For user selection purposes, at BS $b$ the best set $\mathcal{G}_{l}$ should meet two conditions: 1) $\mathbf{h}_{bk_{b}}$ is quasi-orthogonal to $\mathcal{V}_{bk_{b}}$ (similar to DZF), and 2) the elements in $\tilde{\mathbf{H}}_{bk_{b}}(\mathcal{G}_{l})$ are quasi-orthogonal.
If $\epsilon =0$ (interference limited scenario) a strategy for user selection based only on local CSI is hard to define because the channels of all user in $\mathcal{G}_{l}$ are coupled in the SINR expression (\ref{eq:instantaneous_sinr}). In other words, accurate user selection in such scenario requires CSI exchange between BSs.

\section{Metrics of Spatial Compatibility}
\label{sec:metrics_spatial_compatibility}

In this section we answer the question: \textit{what kind of information can be extracted from the local CSI and sent to the CU in order to perform scheduling?}. We define channel metrics whose objective is to measure spatial compatibility between users taking into account the SNR regime and $N_t$.

\subsection{Power Limited Scenario: $N_t \geq B$}
\label{sub_sec:nt_geq_b}

Due to the fact that global CSI is not available at the CU, centralized user selection (e.g., \cite{Niu2013,Marsch2011,Khoshnevis2013}) cannot be performed. In order to design semi-distributed user selection we need to define the type of scheduling control information exchanged between the BSs and the CU.
We say that metric $g_{bl}$ is a function of the local CSI $\mathbf{H}_{b}^{(l)}$ so that $g_{bl}:\mathbb{C}^{N_{t} \times B} \mapsto \mathbb{R}_{+}$. Such mapping computes an approximation of $|\mathbf{h}_{bk_{b}}^{H}\mathbf{w}_{b}|^2$, i.e., it quantifies how profitable is to select the set $\mathcal{G}_{l}$ for transmission at the $b$th BS.
Let $\mathbf{P}_{\mathbf{h}_{bk_{b}}} = \tilde{\mathbf{H}}_{bk_{b}}(\mathcal{G}_{l}) (\tilde{\mathbf{H}}_{bk_{b}}^{H}(\mathcal{G}_{l})\tilde{\mathbf{H}}_{bk_{b}}(\mathcal{G}_{l}))^{-1} \tilde{\mathbf{H}}_{bk_{b}}^{H}(\mathcal{G}_{l})$ be the projector matrix onto $\mathcal{V}_{bk_{b}}$, and $\mathbf{Q}_{\mathbf{h}_{bk_{b}}} = \mathbf{I}_{N_t} -  \mathbf{P}_{\mathbf{h}_{bk_{b}}}$ the projector matrix onto the orthogonal complement of $\mathcal{V}_{bk_{b}}$ \cite{Yanai2011}.
The proposed metric to estimate $|\mathbf{h}_{bk_{b}}^{H}\mathbf{w}_{b}|^2$ is given by:
\begin{equation} \label{eq:metric_of_selection}
g_{bl} = \|\mathbf{Q}_{\mathbf{h}_{bk_{b}}} \mathbf{h}_{bk_{b}}\|^2  + \alpha_{bk_{b}} \|\mathbf{P}_{\mathbf{h}_{bk_{b}}} \mathbf{h}_{bk_{b}}\|^2,
\end{equation}
where $\alpha_{bk_{b}}$ is a function of the type of precoding scheme.

For DZF $\alpha_{bk_{b}} = 0$ for all $\rho_{b}$ since the precoder takes the form $\mathbf{w}_{b}^{(DZF)} = \mathbf{Q}_{\mathbf{h}_{bk_{b}}} \mathbf{h}_{bk_{b}} /\|\mathbf{Q}_{\mathbf{h}_{bk_{b}}} \mathbf{h}_{bk_{b}}\|$ which is the direction of the projection of $\mathbf{h}_{bk_{b}}$ onto $Sp(\tilde{\mathbf{H}}_{bk_{b}}(\mathcal{G}_{l}))^{\perp}$.
For the case of DVSINR, $\mathbf{w}_{b}^{(DVSINR)} = \mathbf{h}_{bk_{b}} /\|\mathbf{h}_{bk_{b}}\|$ as $\rho_{b}\rightarrow 0$, i.e., the precoder is given by the matched filter and we must have $\alpha_{bk_{b}} = 1$ in order to meet $g_{bl}=|\mathbf{h}_{bk_{b}}^{H}\mathbf{w}_{b}^{(DVSINR)}|^2=\|\mathbf{h}_{bk_{b}}\|^2$. When $\rho_{b}\rightarrow\infty$ the precoder is given by $\mathbf{w}_{b}^{(DVSINR)} = \mathbf{Q}_{\mathbf{h}_{bk_{b}}} \mathbf{h}_{bk_{b}} /\|\mathbf{Q}_{\mathbf{h}_{bk_{b}}} \mathbf{h}_{bk_{b}}\|$ and we must have $\alpha_{bk_{b}} = 0$ so that $g_{bl}=|\mathbf{h}_{bk_{b}}^{H}\mathbf{w}_{b}^{(DVSINR)}|^2=\|\mathbf{Q}_{\mathbf{h}_{bk_{b}}} \mathbf{h}_{bk_{b}}\|^2$.
Therefore, $\alpha_{bk_{b}}$ must change it value depending on the SNR regime and the characteristics of the i.i.d. vectors in $\tilde{\mathbf{H}}_{bk_{b}}(\mathcal{G}_{l})$.

\begin{proposition}[]\label{lemma:1}
    The instantaneous effective channel gain $|\mathbf{h}_{bk_{b}}^{H}\mathbf{w}_{b}^{(DVSINR)}|^2$ is given by a nonlinear combination of the orthonormal basis of both $Sp(\tilde{\mathbf{H}}_{bk_{b}}(\mathcal{G}_{l}))$ and $Sp(\tilde{\mathbf{H}}_{bk_{b}}(\mathcal{G}_{l}))^{\perp}$. The metric (\ref{eq:metric_of_selection}) is an approximation of $|\mathbf{h}_{bk_{b}}^{H}\mathbf{w}_{b}^{(DVSINR)}|^2$ and a heuristic definition of the weight $\alpha_{bk_{b}}$ is given by:
    \begin{equation}  \label{eq:heuristic_alpha_dvsinr}
        \alpha_{bk_{b}}=\frac{1}{(\rho_{b}\lambda_{\max}( \tilde{\mathbf{H}}_{bk_{b}}(\mathcal{G}_{l})^{H} \tilde{\mathbf{H}}_{bk_{b}}(\mathcal{G}_{l}) ) + 1)^2}.
    \end{equation}
\end{proposition}
\begin{IEEEproof}
See Appendix~\ref{appendix:4}
\end{IEEEproof}

\subsection{Interference Limited Scenario: $N_t < B$}
\label{sub_sec:nt_less_b}

In this scenario $\epsilon=0$, DZF is not defined \cite{Bjornson2010}, and DVSINR precoding can be implemented but inter-user interference is unavoidable. Moreover, metric (\ref{eq:metric_of_selection}) does not provide information for user selection or cannot be computed. If $B-N_t=1$ then $\mathbf{P}_{\mathbf{h}_{bk_{b}}} = \mathbf{I}_{N_t}$ and we cannot extract useful information from (\ref{eq:metric_of_selection}). If $B-N_t>1$ the matrix $\tilde{\mathbf{H}}_{bk_{b}}^{H}(\mathcal{G}_{l})\tilde{\mathbf{H}}_{bk_{b}}(\mathcal{G}_{l})$ is ill-conditioned\footnote{Observe that $\tilde{\mathbf{H}}_{bk_{b}}^{H}(\mathcal{G}_{l})\tilde{\mathbf{H}}_{bk_{b}}(\mathcal{G}_{l})$ is a matrix of size $B-1 \times B-1$ which has $N_t$ non-zero eigenvalues. When $B-N_t>1$ the ratio $\lambda_{\max}(\tilde{\mathbf{H}}_{bk_{b}}^{H}(\mathcal{G}_{l})\tilde{\mathbf{H}}_{bk_{b}}(\mathcal{G}_{l}))/ \lambda_{\min}(\tilde{\mathbf{H}}_{bk_{b}}^{H}(\mathcal{G}_{l})\tilde{\mathbf{H}}_{bk_{b}}(\mathcal{G}_{l})) \rightarrow\infty$ and the matrix is close to singular \cite{Gentle2007}.} and $\mathbf{P}_{\mathbf{h}_{bk_{b}}}$ is no longer a projector matrix.
Therefore, we want to define a metric of the form $g_{bl} = \|\mathbf{h}_{bk_{b}}\|^2 f(\mathbf{H}_{b}^{(l)}, \rho_{b})$. The function $f(\mathbf{H}_{b}^{(l)}, \rho_{b})$ must become 1 as $\rho_{b} \rightarrow 0$ whilst its value should change according to the strength of $\mathbf{h}_{bk_{b}}$ and its spatial relation with $\mathcal{V}_{bk_{b}}$ as $\rho_{b} \rightarrow \infty$.

Consider two grouped users $k_{b} \in \{ \mathcal{S}_{b} \cap \mathcal{G}_{l} \}$, $k_{j} \in \{ \mathcal{S}_{j} \cap \mathcal{G}_{l} \}$, and define $m_{b}^{(k_{b}k_{j})} = \|\mathbf{h}_{bk_{b}}\|^2/\|\mathbf{h}_{bk_{j}}\|^2$. $m_{b}^{(k_{b}k_{j})}$ can provide a coarse estimation of the location of the users regarding the $b$th BS.  If $m_{b}^{(k_{b}k_{j})} \approx 1$ this may suggest either that $k_{b}$ and $k_{j}$ are close to each other at the cell-edge, or that $k_{b}$ is far from BS $b$ and transmission over channel $\mathbf{h}_{bk_{b}}$ could be affected by strong interference. For $m_{b}^{(k_{b}k_{j})} \gg 1$, $k_{b}$ may be close to BS $b$ or $k_{j}$ is either far from BS $b$ or experiencing deep fading. If $m_{b}^{(k_{b}k_{j})} < 1$ fading is large in $\mathbf{h}_{bk_{b}}$ and transmission may be infeasible. In order to quantify how strong and reliable for transmission is $\mathbf{h}_{bk_{b}}$ using local CSI, define the coefficient $M_{bk_{b}}^{(l)}$ as:

\begin{equation} \label{eq:metric_2_component_M}
    M_{bk_{b}}^{(l)} = \frac{\|\mathbf{h}_{bk_{b}}\|^2}{ \prod_{j=1,j\neq b}^{B} \|\mathbf{h}_{bk_{j}}\|^{2/(B-1)}  }, 
\end{equation}
where the denominator is the geometric mean of the squared norms of the column vectors of $\tilde{\mathbf{H}}_{bk_{b}}(\mathcal{G}_{l})$. Using the geometric mean has two objectives: collecting in a single quantity the strength of the channels $\{\mathbf{h}_{bk_{j}}\}_{k_{j} \in \mathcal{G}_{l} \setminus \{k_{b}\}}$ and considering the effects of each magnitude equally\footnote{This is not the case for the arithmetic mean since the magnitudes $\{\|\mathbf{h}_{bk_{j}}\|^2\}_{k_{j} \in \mathcal{G}_{l} \setminus \{k_{b}\}}$ may have a large variance in which case the smallest magnitudes would be neglected.} in the averaging operation.

We also need to estimate the spatial compatibility between all the elements of $\mathbf{H}_{b}^{(l)}$, the degradation due to correlation in $\tilde{\mathbf{H}}_{bk_{b}}(\mathcal{G}_{l})$ and the effects of $\rho_{b}$. In other words, we need an operation similar to the NSP. Define the metric for spatial compatibility as
\begin{equation} \label{eq:metric_2_component_C}
    \zeta_{bk_{b}}^{(l)} = \frac{\left|\det \left( (\mathbf{H}_{b}^{(l)})^{H}\mathbf{H}_{b}^{(l)} \right)\right| }{ \left|\det\left( \rho_{b}^{-1}\mathbf{I}_{N_{t}} + \tilde{\mathbf{H}}_{bk_{b}}(\mathcal{G}_{l})(\tilde{\mathbf{H}}_{bk_{b}}(\mathcal{G}_{l}))^{H} \right)\right| },
\end{equation}
which is the ratio between the volume of a $B \times B$ matrix over the the volume of a $N_t \times N_t$ matrix. Recall that the determinant measures the volume spanned by the columns of a matrix. The more orthogonal the column vectors of a matrix, the larger the value of its determinant \cite{Gentle2007}.

The heuristic metric for user selection is defined as
\begin{equation} \label{eq:metric_of_selection_2}
g_{bl} = \|\mathbf{h}_{bk_{b}}\|^2  \left(\alpha_{bk_{b}}   + (1-\alpha_{bk_{b}}) M_{bk_{b}}^{(l)} \zeta_{bk_{b}}^{(l)} \right),
\end{equation}
where $\alpha_{bk_{b}}$ is given by (\ref{eq:heuristic_alpha_dvsinr}). Observe that in the low SNR $\alpha_{bk_{b}}\rightarrow 1$ which yields $g_{bl} \approx \|\mathbf{h}_{bk_{b}}\|^2$. In the high SNR $\alpha_{bk_{b}}\rightarrow 0$ and the selection metric is $g_{bl} \approx \|\mathbf{h}_{bk_{b}}\|^2 M_{bk_{b}}^{(l)} \zeta_{bk_{b}}^{(l)}$.

\subsection{NSP Approximation}
\label{sub_sub_sec:approximation_nsp}
The NSP operation $\|\mathbf{Q}_{\mathbf{h}_{bk_{b}}} \mathbf{h}_{bk_{b}}\|^2 = \|\mathbf{h}_{bk_{b}}\|^2 \sin^2 \theta_{\mathcal{V}_{bk_{b}}\mathbf{h}_{bk_{b}}} $ can be approximated using the inner products of the elements of $\mathbf{H}_{b}^{(l)}$ which reduces the number of arithmetic operations required to compute metrics (\ref{eq:metric_of_selection}) or (\ref{eq:metric_of_selection_2}). The term $\theta_{\mathcal{V}_{bk_{b}}\mathbf{h}_{bk_{b}}}$ is the angle between $\mathbf{h}_{bk_{b}}$ and the subspace $\mathcal{V}_{bk_{b}}$. For two i.i.d. channels $\mathbf{h}_{bk_{b}}$ and $\mathbf{h}_{bi_{b}}$ at the $b$th BS, the spatial compatibility between them can be measured by the coefficient of correlation defined as \cite{Yanai2011}:
\begin{equation}\label{eq:coefficient_correlation}
    \eta_{\mathbf{h}_{bk_{b}} \mathbf{h}_{bi_{b}}}  =  \frac{|\left< \mathbf{h}_{bk_{b}}, \mathbf{h}_{bi_{b}} \right>|}{ \| \mathbf{h}_{bk_{b}} \| \| \mathbf{h}_{bi_{b}} \|},
\end{equation}
where the coefficient $0\leq \eta_{\mathbf{h}_{bk_{b}} \mathbf{h}_{bi_{b}}} = \cos \theta_{\mathbf{h}_{bk_{b}} \mathbf{h}_{bi_{b}}} \leq 1$ geometrically represents the cosine of the angle between the two channel vectors.
The coefficient $\sin^2 \theta_{\mathcal{V}_{bk_{b}}\mathbf{h}_{bk_{b}}}$ that scales $\|\mathbf{h}_{bk_{b}}\|^2$ in a NSP operation can be computed as \cite{Yanai2011}:
\begin{equation}\label{eq:coefficient_determination_from_partial_correlation_coefficients}
    \sin^2 \theta_{\mathcal{V}_{bk_{b}}\mathbf{h}_{bk_{b}}} = (1-\eta_{\mathbf{h}_{bk_{b}}\pi(1)}^{2}) \ldots (1-\eta_{\mathbf{h}_{bk_{b}}\pi(i)|\pi(1)\ldots\pi(i-1)}^{2}),
\end{equation}
where $\pi(i)$ is the $i$th ordered element of $\tilde{\mathbf{H}}_{bk_{b}}(\mathcal{G}_{l})$ and $\eta_{\mathbf{h}_{bk_{b}}\pi(i)|\pi(1)\ldots\pi(i-1)}$ is the partial correlation coefficient between the channel vector $\mathbf{h}_{bk_{b}}$ and the selected vector associated with $\pi(i)$ eliminating the effects due to the previous ordered elements $\pi(1),\pi(2),\ldots,\pi(i-1)$.

If the correlation coefficients (\ref{eq:coefficient_correlation}) are used instead of the partial correlation coefficients in (\ref{eq:coefficient_determination_from_partial_correlation_coefficients}) a suboptimal evaluation of $\sin^2 \theta_{\mathcal{V}_{bk_{b}}\mathbf{h}_{bk_{b}}}$ can be computed.
Using this approximation of the NSP, the reported metric to the CU by the $b$th BS for the user $k_{b} \in \{ \mathcal{S}_{b} \cap \mathcal{G}_{l} \}$ is given by:

\begin{equation} \label{eq:approximation_coefficient_determination}
g_{bl} = \|\mathbf{h}_{bk_{b}}\|^2 \prod_{i_{b}\neq k_{b}, i_{b} \in \mathcal{G}_{l}} \sin^{2}\theta_{\mathbf{h}_{bk_{b}} \mathbf{h}_{bi_{b}}} .
\end{equation}

Observe that metric (\ref{eq:approximation_coefficient_determination}) can be computed even if $N_t < B$ since (\ref{eq:coefficient_correlation}) is independent of $B$ and exists for all $N_t \geq 2$.
If $N_t \geq B$ metric (\ref{eq:approximation_coefficient_determination}) is an upper bound of the NSP. This means that $\|\mathbf{h}_{bk_{b}}\|^{2}$ is scaled by a coefficient larger than $\sin^2 \theta_{\mathcal{V}_{bk_{b}}\mathbf{h}_{bk_{b}}}$ which prioritizes the channel magnitude over the spatial compatibility when the user selection is performed. The relationship between the real and the approximated expected values of the NSP is presented in the following proposition.

\begin{proposition}[]\label{prop:over_approximation_nsp}
    For $N_t \geq B$ it holds that the average value of  metric (\ref{eq:approximation_coefficient_determination}) is an upper bound of the average metric (\ref{eq:metric_of_selection}) with $\alpha_{bk_{b}} = 0$, i.e., the NSP, so that:
    \begin{equation}\label{eq:upper_bound_app_nsp}
        \mathbb{E}\left[ \|\mathbf{h}_{bk_{b}} \mathbf{Q}_{\mathbf{h}_{bk_{b}}}\|^2 \right] \leq \mathbb{E}\left[ \|\mathbf{h}_{bk_{b}}\|^2  \prod_{i_{b}\neq k_{b}, i_{b} \in \mathcal{G}_{l}} \sin^{2}\theta_{\mathbf{h}_{bk_{b}} \mathbf{h}_{bi_{b}}}  \right]
    \end{equation}
\end{proposition}
\begin{IEEEproof}
See Appendix~\ref{appendix:4}
\end{IEEEproof}

\section{The Multicell User Selection}
\label{sec:multicell_user_selection}

Once that $g_{bl}$ has extracted spatial compatibility information from the multiuser channel matrix $\mathbf{H}_{b}^{(l)}$ we need to answer two questions: 1) \textit{what should be the optimization over the metrics $g_{bl}$ at the CU in order to find the set $\mathcal{G}_{l^{*}}$ of most spatially compatible users?} and 2) \textit{how to minimize the number of metrics $g_{bl}$ computed per BS so that $\mathcal{G}_{l^{*}}$ achieves close-to-optimal performance and  multiuser diversity is preserved?}

\subsection{Exhaustive Search Selection over the Metrics}
\label{sub_sec:optimum_selection}

The optimal solution of (\ref{eq:sum_rate_max_problem}) can be only found by exhaustive searching over the achievable rates of the sets $\mathcal{G}_{l} \ \forall l \in \{1, \ldots, L\}$. Such a search requires global CSI at the CU and the computation of $BL$ precoders in order to accurately evaluate the $L$ possible achievable sum rates.
A sub-optimal solution to (\ref{eq:sum_rate_max_problem}) can be found by avoiding the full CSI exchange with the CU and instead reporting the metrics computed by (\ref{eq:metric_of_selection}), (\ref{eq:metric_of_selection_2}), or (\ref{eq:approximation_coefficient_determination}).
Assuming that all BSs know the $L$ ordered sets, the $b$th BS computes the metrics $g_{bl}$ $\forall l$ and report them to the CU where the index of the set that is chosen to perform coordinated transmission is found solving the following problem:

\begin{equation}\label{eq:best_metric_group_selection_problem}
    l^{\star} =  \arg \underset{l \in \{1,\dots,L\} }{\max} \prod_{b=1}^{B} g_{bl}.
\end{equation}

Bearing in mind that $g_{bl}$ attempts to estimate the effective channel gains, the rationale behind the product in (\ref{eq:best_metric_group_selection_problem}) is that for MISO transmission a set of users maximizing the product of their effective channel gains also achieves maximum sum rate \cite{Chan2007}.
In our scenario, taking the product of the metrics assigns the same priority to each independent metric $g_{bl}$ $\forall b$. This means that the computation of $l^{\star}$ is not biased by a dominant metric $g_{bl}\gg g_{jl}$ $\forall j \neq b$ for a given set $l$, which would be only beneficial to BS $b$.
Once that $l^{\star}$ has been found, the BSs use the matrices $\mathbf{H}_{b}^{(l^{\star})}$ $\forall b$ to locally compute the precoders which are used to sub-optimally solve (\ref{eq:sum_rate_max_problem}).

\subsection{Search Space Pruning}
\label{sub_sec:search_space_pruning}

Previously we have discussed that solving (\ref{eq:best_metric_group_selection_problem}) does not require global CSI but $L$ metrics are reported from each BS to the CU. If the number of cell-edge users is large ($|\mathcal{S}| \gg BN_t$) computing the metrics for all user permutations $L$ may become prohibitive. In single-cell systems the authors in \cite{Sharif2007} showed that for fixed $N_t$ and single-antenna users, the system capacity under spatial division multiple access scales by $N_t\log(\log(|\mathcal{S}_{b}|))$ at the $b$th BS. This result means that multiuser diversity provides a marginal contribution to the capacity enhancement unless $|\mathcal{S}_{b}|\rightarrow \infty$. Similar conclusions extend to multi-cell systems operating in JT mode (e.g., \cite{Huh2012,Khoshnevis2013}). Numerical results in \cite{Huh2012} show that multiuser diversity is beneficial for BS cooperation when only a fraction of the total number of users is considered to participate in the selection process.
For a multi-cell JT system employing ZF precoding \cite{Khoshnevis2013}, $BN_t$ transmit antennas can serve at most the same number of single-antenna users, and low-complexity user selection algorithms can be extended from single-cell systems \cite{Tu2003,Yoo2006,Chan2007,Fuchs2007,Castaneda2014}.

In our CBF scenario we want: to achieve multiplexing gain; decrease the solution space's size of problem (\ref{eq:best_metric_group_selection_problem}) by selecting a small fraction of competing users from $\mathcal{S}$; and preserve multiuser diversity when selecting the competing users.
In order to find a subset $\hat{\mathcal{S}}_{b}\subseteq \mathcal{S}_{b}$ at BS $b$, let $\mathbf{h}_{bk_{b}} = [{h}_{bk_{b}1}, \ldots,  {h}_{bk_{b}N_t}]^T$ be the channel of the user $k_{b} \in \mathcal{S}_{b}$ where ${h}_{bk_{b}n}$ is the channel component of the $n$th antenna.
Consider the following:
1) For DZF efficient user selection must be focused on finding quasi-orthogonal users regarding the SNR regime.
2) For DVSINR efficient user selection in the low SNR is determined by the channel magnitude. 3) In the high SNR the effective channel gains of DZF and DVSINR are similar and efficient user selection must find spatially orthogonal users.
A fast way to find a set of quasi-orthogonal users in JT systems is by applying a ranking-based per-antenna selection as in \cite{Khoshnevis2013}. The idea behind such selection is that for two user, $k_{b}$ and $i_{b}$ having $|{h}_{bk_{b}n}| > |{h}_{bk_{b}n'}|$ $\forall n'\neq n$, $|{h}_{bi_{b}m}| > |{h}_{bi_{b}m'}|$ $\forall m'\neq m$, and $\forall n\neq m$, the inner product $\left< \mathbf{h}_{bk_{b}}, \mathbf{h}_{bi_{b}} \right>$ decreases as the magnitude of each dominant antenna $n$ and $m$ increases, i.e., they become quasi-orthogonal.

In our scenario we require the channel of the selected user $k_{b} \in \{ \mathcal{S}_{b} \cap \mathcal{G}_{l} \}$ of BS $b$ to be as orthogonal as possible w.r.t. the channels in $\tilde{\mathbf{H}}_{bk_{b}}(\mathcal{G}_{l})$. Therefore, the per-antenna ranking can be used for pre-selecting the users with maximum per-antenna channel magnitude. In this way a user $k_{b} \in \hat{\mathcal{S}}_{b}$ will have a dominant antenna (spatial direction) $n$ and it is likely that channels in $\tilde{\mathbf{H}}_{bk_{b}}(\mathcal{G}_{l})$ do not have per-antenna channel magnitudes similar or closed to $|{h}_{bk_{b}n}|$ at the same antenna $n$ due to path-loss effects, which provides certain degree of spatial compatibility.

Define the dominant user for the antenna $n$ at BS $b$ as
\begin{equation} \label{eq:donimant_user_per_antenna}
    k_{b(n)} = \arg \underset{ i_{b} \in \mathcal{S}_{b} }{ \max  } \ \ |{h}_{bi_{b}n}|,
\end{equation}
and let the user with the largest channel magnitude be
\begin{equation} \label{eq:donimant_user_per_all_antennas}
    k_{b(\max)} = \arg \underset{ i_{b} \in \mathcal{S}_{b} }{ \max  } \ \ \|\mathbf{h}_{bi_{b}}\|,
\end{equation}
where the subset of users that will participate in the selection process at the $b$th BS is defined as
\begin{equation} \label{eq:set_preselected_users}
    \hat{\mathcal{S}}_{b} = \{ k_{b(n)} \}_{n=1}^{N_t} \cup \{ k_{b(\max)} \}.
\end{equation}

\begin{figure}[!t]
  \centering
  \includegraphics[width=3.4in]{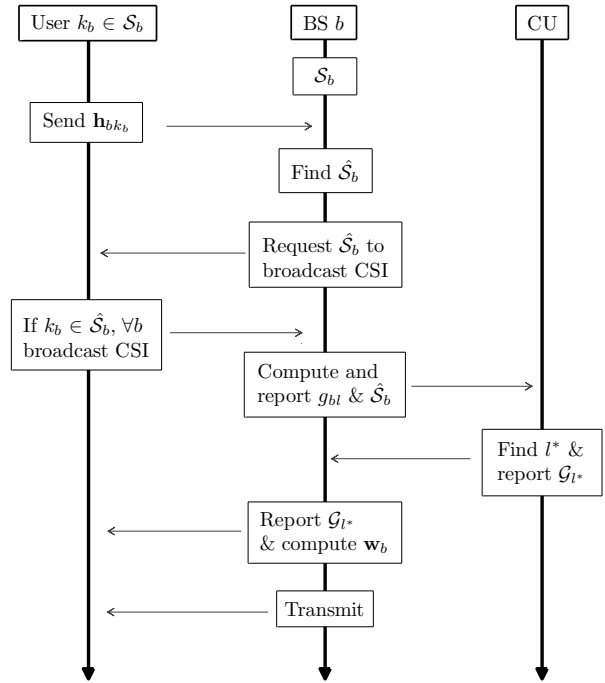}\\
  \caption{Proposed Coordinate Scheduling and CBF Transmission}
  \label{fig:protocol}
\end{figure}

This user pre-selection reduces the size of the search space because it only considers the strongest users per spatial direction per BS. Including $k_{b(\max)}$ in the set $\hat{\mathcal{S}}_{b}$ guarantees that for DVSINR the strongest user will be considered for selection. Observe that the index $k_{b(\max)}$ can be one or more of the indices $k_{b(n)}$ $\forall n$ which may be repeated as well, and $|\hat{\mathcal{S}}_{b}|$ can be at most $N_t + 1$. Assuming that $|\mathcal{S}_{b}| \geq N_t+1$, $\forall b$ the minimum number of reported metrics per BS, denoted as $L_{r}$, that will be used to solve (\ref{eq:best_metric_group_selection_problem}) is bounded as follows:

\begin{equation} \label{eq:bounds_number_reported_metrics}
    L_{r} = \prod_{b=1}^{B} |\hat{\mathcal{S}}_{b}| \leq  (N_t + 1)^{B} \leq L = \prod_{b=1}^{B} |\mathcal{S}_{b}|
\end{equation}
and notice that $L_{r}$ is independent of $|\mathcal{S}_{b}|$ $\forall b$. Fig.~\ref{fig:protocol} illustrates the sequence of information exchange between users, BSs, and the CU considering channel metrics and search space pruning.

\section{Numerical Results}
\label{sec:numerical_results}

In this section the performance of the joint distributed linear precoding and user selection is illustrated numerically. The results are obtained using the deployment described in Section~\ref{sec:system_model} with $B=3$, cell radius $r=1000$(m), and cell-edge cooperation area of radius $r_{coop}=300$(m). For simplicity all BSs have the same number of users $K$. The long-term channel power gain is proportional to $1/d_{bk_{b}}^{4}$ where $d_{bk_{b}}$ is the distance between user $k_{b}$ and BS $b$ given in meters. We assume perfect CSI at each BS, the average sum rate is given in (bps/Hz), and the results are averaged over 10,000 channel realizations.
The results are computed by assigning $P_{b}=P$ for all $b \in \{1,\ldots,B\}$ and the same SNR regime at the cell border to all BSs, i.e., $\rho = P/\sigma_{n}^{2}$.

The system performance benchmark is given by the optimal solution of problem (\ref{eq:sum_rate_max_problem}) which is achieved by global CSI at the CU and is referred to as O-GCSI.
In order to solve problem (\ref{eq:best_metric_group_selection_problem}) two strategies are implemented: 1) considering all $L$ user permutations, and 2) applying the search space pruning with $L_r$ user permutations. For scenarios where $N_t \geq B$, the results obtained for (\ref{eq:metric_of_selection}) are referred to as O-MUS (metric of user selection) when $L$ is considered, or R-MUS if $L_r$ is used. Similarly, metric (\ref{eq:approximation_coefficient_determination}) is referred to as O-NSPA (NSP approximation) for $L$ and R-NSPA for $L_r$. If $N_t < B$ the results for (\ref{eq:metric_of_selection_2}) are referred to as O-MUS2 and R-MUS2 for $L$ and $L_r$ respectively.
In order to highlight how the proposed metrics exploit multiuser diversity we compare their performance w.r.t. a selfish user selection where each BS transmits to its strongest user (maximum channel norm) referred to as Max-SNR.

\subsection{Sum rate vs SNR ($\rho$)}
\label{sub_sec:rate_vs_snr}

\begin{figure}[!t]
  \centering
  \includegraphics[width=3.4in]{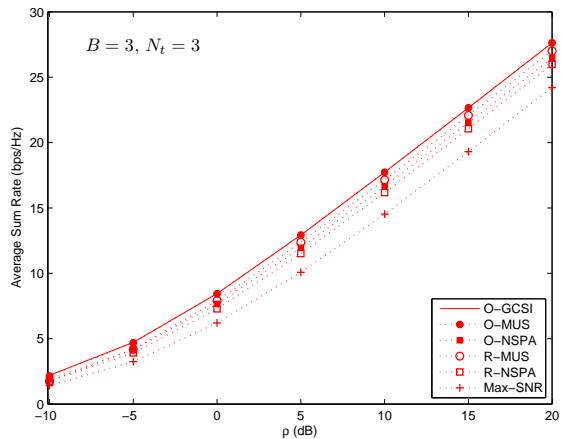}\\ 
  \caption{Average Sum Rate as a function of $\rho$(dB) for DZF precoding with $K=10$, $B=3$ and $N_t=3$.}
  \label{fig:capacity_vs_snr_B_3_N_3_K_10_DZF}
\end{figure}

\begin{figure}[!t]
  \centering
  \includegraphics[width=3.4in]{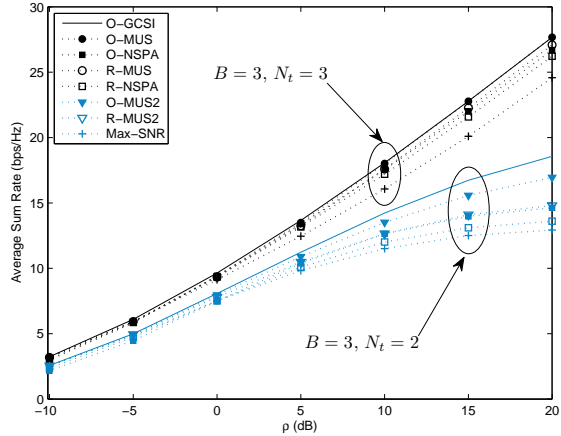}\\ 
  \caption{Average Sum Rate as a function of $\rho$(dB) for DVSINR precoding with $K=10$, $B=3$ and $N_t \in \{2, 3\}$.}
  \label{fig:capacity_vs_snr_B_3_N_3_K_10_DVSINR}
\end{figure}

The average sum rate as function of $\rho$ (dB) for DZF and DVSINR is displayed in Fig.~\ref{fig:capacity_vs_snr_B_3_N_3_K_10_DZF} and Fig.~\ref{fig:capacity_vs_snr_B_3_N_3_K_10_DVSINR} respectively.
In Fig.~\ref{fig:capacity_vs_snr_B_3_N_3_K_10_DZF} for the case $N_t = 3$, $B = 3$, and a target rate of 13(bps/Hz) the O-NSPA requires about 1(dB) extra to achieve the target compared to O-GCSI. For a target $\rho$ of 10(dB) the O-NSPA has a gap about 1(bps/Hz) compared to O-MUS. The simulated scenarios considered $K=10$ users per BS, O-GCSI, O-MUS, O-NSPA require to evaluate $L=10^3$ metrics per BS while R-MUS, R-NSPA require $L_r \leq 4^3$ for $N_t=3$. For $\rho=10$(dB) R-MUS and R-NSPA achieve 96\% and 91\% of the optimal performance O-GCSI, which shows the effectiveness of the search space pruning for CBF systems under DZF precoding. The performance gap between R-MUS and R-NSPA w.r.t. O-GCSI is about 1\% and 2.5\% respectively  for $N_t=4$ and $\rho=10$(dB).

Fig.~\ref{fig:capacity_vs_snr_B_3_N_3_K_10_DVSINR} show results for DVSINR with $B=3$ and $N_t \in \{2, 3\}$. In the case $N_t\geq B$ the results for O-MUS and O-GCSI are quite closed and the sum rate of O-MUS is sub-optimal in the middle SNR range. The performance gap of O-MUS is less than 3\% in the SNR range $\rho \in [-10,10]$ and such a gap vanishes for other values of $\rho$. In the case of O-NSPA for $N_t\geq B$, it achieves up to 96\% of the rate of O-GCSI in the whole SNR range. For $N_t=3$ and $\rho=10$(dB), the performance gap between R-MUS and R-NSPA w.r.t. O-GCSI is about 3\% and 5\% respectively.

We have discussed that for the interference limited scenario ($N_t < B$), the SINR (\ref{eq:instantaneous_sinr}) of user $k_{b} \in \mathcal{S}_{b}$ depends on all its cross-channels $\{\mathbf{h}_{jk_{b}}\}_{j=1,j\neq b}^{B}$. However, BS $b$ only knows $\mathbf{h}_{bk_{b}}$ and an accurate user selection must take into account both the effective channel gain over the direct and cross channels, unlikely the case $N_t \geq B$. The figure shows that considering all user permutations $L$ for metric (\ref{eq:metric_of_selection_2}), O-MUS2, is highly efficient in the low SNR regime and it achieves up to 91\% of the sum rate of O-GCSI when $\rho=20$(dB). In contrast, O-NSPA cannot exploit multiuser diversity efficiently and only achieves 78\% of the O-GCSI performance at the same SNR. Accounting for the search space pruning, R-NSPA and R-MUS2 attain 79\% and 73\% of the O-GCSI performance, respectively. These results show the effectiveness of the propose metric  (\ref{eq:metric_of_selection_2}) and highlight the fact that we rely on $L$ metrics per BS in order to achieve acceptable performance and compensate the lack of CSI knowledge of other BSs.

It is worth mentioning that the performance in the interference limited scenario ($N_t < B$) can be improved by joint power allocation (coordinated by the CU) but this requires global CSI knowledge of the scheduled users and their precoder vectors. This kind of power optimization is out of the scope of the paper and we refer to \cite{Chiang2008} for an in-depth survey.

\subsection{Sum rate vs $K$ }
\label{sub_sec:rate_vs_dof}

\begin{figure}[!t]
  \centering
  \includegraphics[width=3.4in]{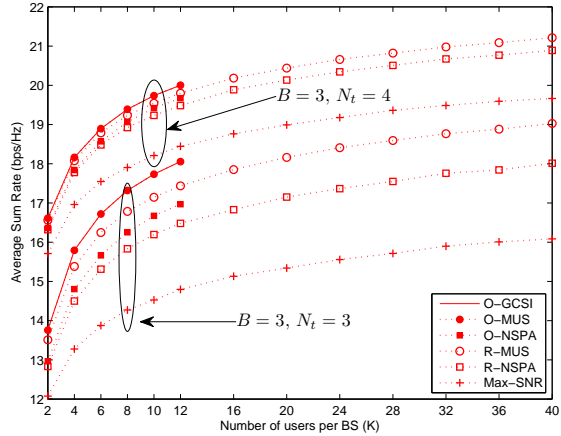}\\
  \caption{Average Sum Rate as a function of the number of users per BS (K) for DZF with $\rho=10$(dB), $B=3$, and $N_t \in \{3, 4\}$.}
  \label{fig:rate_vs_users_B_3_var_N_DZF}
\end{figure}

Fig.~\ref{fig:rate_vs_users_B_3_var_N_DZF} shows the average sum rate as a function of the number of users (multiuser diversity) $K$ for DZF and $\rho=10$(dB). The figure illustrates the average sum rate of two scenarios where $N_t=B$ and $N_t > B$. In our CBF scenario, numerical results show that the set of users maximizing the product of their effective channel gains (computed with local CSI) achieve maximum sum rate for DZF.
For $K=12$ O-MUS and O-GCSI overlap, O-NSPA attains up to 94\% of the sum rate of O-GCSI for $N_t=3$ and 98\% for $N_t=4$.
The performance of R-MUS and R-NSPA illustrates the benefits of the proposed search space pruning. For $K=12$, $N_t=3$, and $B=3$ the sum rate gap between R-MUS and O-MUS is less than 4\% while the gap between R-NSPA and O-NSPA is less than 3\% but the gain in terms of computed metrics per BS is remarkable since $L_r \leq 64 < L=1728$.

\begin{figure}[!t]
  \centering
  \includegraphics[width=3.4in]{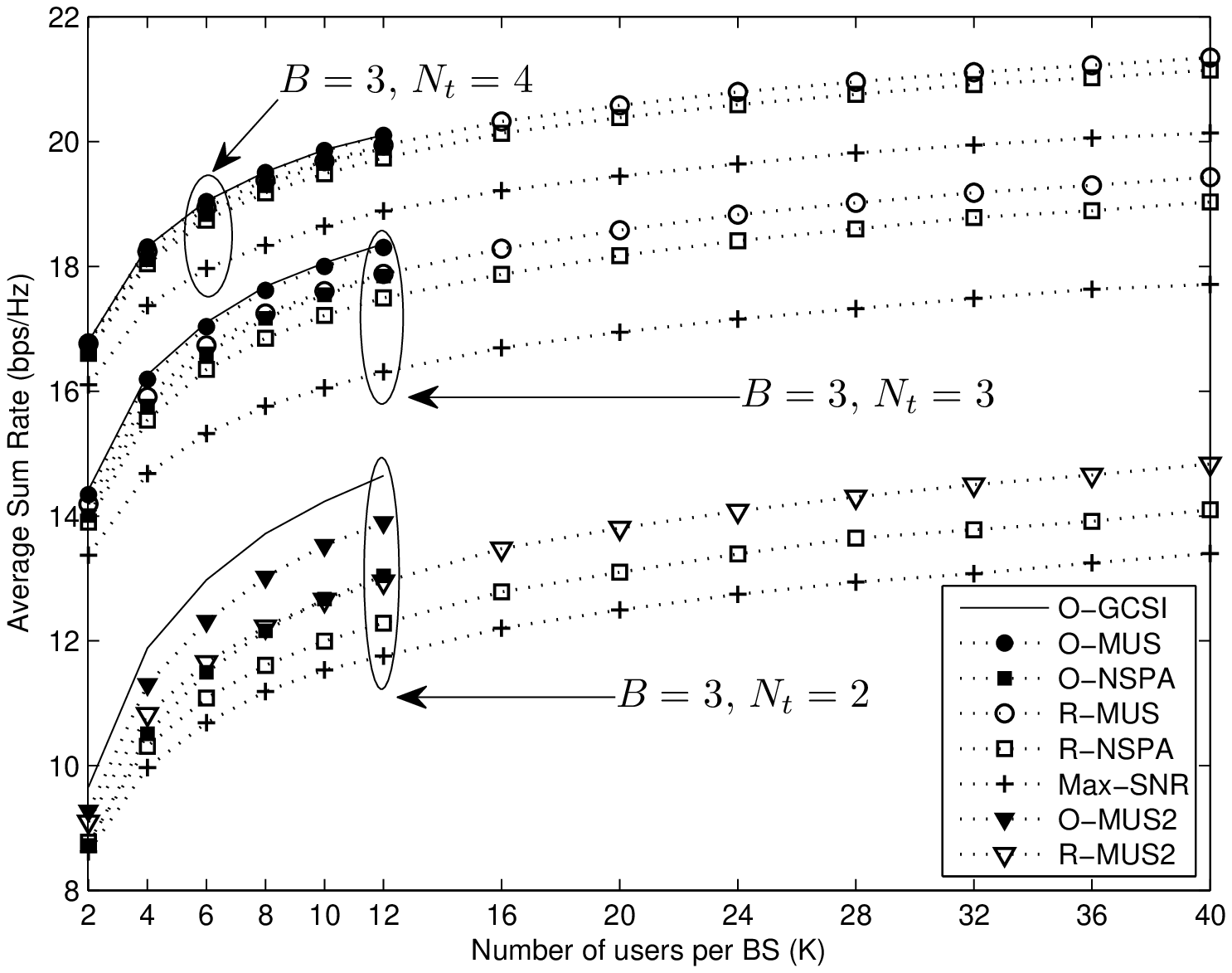}\\
  \caption{Average Sum Rate as a function of the number of users per BS (K) for DVSINR with $\rho=10$(dB), $B=3$, and $N_t \in \{2, 3, 4\}$.}
  \label{fig:rate_vs_users_B_3_var_N_DVSINR}
\end{figure}

Fig.~\ref{fig:rate_vs_users_B_3_var_N_DVSINR} shows the sum rate as a function of $K$ for DVSINR and $\rho=10$(dB). The figure shows the performance for three scenarios with fixed $B=3$ and $N_t \in \{4 ,3, 2\}$. For $N_t=4$ and $K=12$, O-MUS and O-NSPA achieve 99\% and 98\% respectively of the benchmark sum rate. For $N_t=3$ and $K=12$, O-MUS and O-NSPA achieve 99\% and 97\% of the O-GCSI performance respectively. When $N_t=2$, the O-MUS2 and O-NSPA achieve 94\% and 89\% of the optimal sum rate, respectively.
The computational gains due to search space pruning are larger when $N_t \geq B$ and slightly reduced when $N_t<B$. For the latter scenario the performance gap between O-MUS2 and R-MUS2 is about 7\% while the gap between O-NSPA and R-NSPA is about 6\% for $K=12$.

The performance gap between O-MUS and O-NSPA for both DZF in Fig.~\ref{fig:rate_vs_users_B_3_var_N_DZF} and DVSINR in Fig.~\ref{fig:rate_vs_users_B_3_var_N_DVSINR} reduces considerably by adding one extra antenna per BS. These results suggest that metric (\ref{eq:approximation_coefficient_determination}) should be preferred instead of (\ref{eq:metric_of_selection}) for system where $N_t \gg B$. The advantage of (\ref{eq:approximation_coefficient_determination}) is that it only requires inner product and vector norm operations while (\ref{eq:metric_of_selection}) requires several matrix operations.
The results illustrate the effectiveness of the proposed metrics, two of them capturing more accurately spatial compatibility of the multiuser channels (\ref{eq:metric_of_selection}) and (\ref{eq:metric_of_selection_2}), and metric (\ref{eq:approximation_coefficient_determination}) that is less computationally demanding and independent of the relation between $B$ and $N_t$.
The advantage of using the search space pruning is explicit if $K > N_t$. For $K=40$ there exists $L=40^{B}$ possible metrics per BS, however, the CU requires at most $(N_t+1)^{B}$ metrics per BS in order to perform efficient user selection.

\section{Conclusions}
\label{sec:conslusions}
In this paper we have addressed the sum rate maximization problem for multi-cell systems in the CBF transmission mode with limited message exchange between BSs.
Considering that CSI is not globally available we adopt two distributed linear precoding schemes with defined structures, DZF and DVSINR, and discussed the characteristics of their respective effective channel gains. We showed that user selection must be based on the precoding technique that is implemented and channel metrics were designed for two scenarios 1)$N_t\geq B$ and 2)$N_t<B$. The objective of the metrics is to use the local CSI to provide an estimation of the achievable effective channel gains for each precoder technique. Using the metrics at the CU, we designed an algorithm that selects a set of spatially compatible users. Finally, we proposed a method for search space pruning which dramatically reduces the number of metrics reported from the BSs to the CU and preserves multiuser diversity. Our algorithm and metrics for user selection were assessed by simulations and numerical results show their potential to improve performance in coordinated multi-cell systems with limited message exchange between BSs.

\appendices
\section{}
\label{appendix:1}
\ \ \textit{Proof of the Proposition~\ref{thm:1}:}
In what follows, we slightly abuse the notation and omit the user and BS subindices.
Consider the channel of the served user $\mathbf{h}$, its precoding vector $\mathbf{w}$ defined in (\ref{eq:define_dzf_precoder}), and its aggregate interference matrix $\tilde{\mathbf{H}}$, define $\tilde{\mathbf{V}}=null(\tilde{\mathbf{H}})$ the matrix that contains the orthonormal vectors $\{\tilde{\mathbf{v}}_{i}\}_{i=1}^{\epsilon}$ and $\rho = P/\sigma_{n}^{2}$. The effective channel gain is given by:

\begin{IEEEeqnarray}{rCl}
    |\mathbf{h}^{H}\mathbf{w} |^2
    & = &  |Tr(\mathbf{h}\mathbf{w}^{H})|^{2}  \IEEEyessubnumber \label{eq:eff_chb_gain_dzf_a} \\
    & = & \frac{ |Tr(\tilde{\mathbf{V}}\tilde{\mathbf{V}}^{H}\mathbf{h}\mathbf{h}^{H})|^{2} } { |\mathbf{h}^{H}\tilde{\mathbf{V}}\tilde{\mathbf{V}}^{H}\mathbf{h}| }  \IEEEyessubnumber \label{eq:eff_chb_gain_dzf_b} \\
    & = & |\mathbf{h}^{H}\tilde{\mathbf{V}}\tilde{\mathbf{V}}^{H}\mathbf{h}| \IEEEyessubnumber \label{eq:eff_chb_gain_dzf_c} \\
    & = & \|\tilde{\mathbf{V}}\tilde{\mathbf{V}}^{H}\mathbf{h}\|^2  \IEEEyessubnumber \label{eq:eff_chb_gain_dzf_d} \\
    & = & \| \left(\sum_{i=1}^{\epsilon} \tilde{\mathbf{v}}_{i}\tilde{\mathbf{v}}_{i}^{H}\right) \mathbf{h}\|^2  \IEEEyessubnumber \label{eq:eff_chb_gain_dzf_e} \\
    & = & \|\mathbf{h}\|^2 \sum_{i=1}^{\epsilon} \cos^{2} \theta_{\mathbf{h} \tilde{\mathbf{v}}_{i} }  \IEEEyessubnumber \label{eq:eff_chb_gain_dzf_f}
\end{IEEEeqnarray}
where (\ref{eq:eff_chb_gain_dzf_a}) is due to the fact that $\|\mathbf{w}\|=1$ and $<\mathbf{h},\mathbf{w}> = Tr(\mathbf{h}\mathbf{w}^{H})$. (\ref{eq:eff_chb_gain_dzf_b}) is given by substituting $\mathbf{w}^{H}$ into (\ref{eq:eff_chb_gain_dzf_a}) and properties of the trace and outer product \cite{Gentle2007}.
(\ref{eq:eff_chb_gain_dzf_c}) is done by expressing the denominator in (\ref{eq:eff_chb_gain_dzf_b}) in the form of the numerator.
(\ref{eq:eff_chb_gain_dzf_d}) The basis of the null space projection $\tilde{\mathbf{V}}$ can be used to compute the projection matrix $\mathbf{Q}_{\mathbf{h}} = \tilde{\mathbf{V}}\tilde{\mathbf{V}}^{H}$ \cite[\S 2.6]{Golub1996}. The projector matrix is an idempotent matrix, i.e., $\mathbf{Q}_{\mathbf{h}} = \mathbf{Q}_{\mathbf{h}}\mathbf{Q}_{\mathbf{h}}^{H}= \mathbf{Q}_{\mathbf{h}}^{H}\mathbf{Q}_{\mathbf{h}}$ \cite{Yanai2011}.
(\ref{eq:eff_chb_gain_dzf_e}) is a decomposition of the form $\tilde{\mathbf{V}}\tilde{\mathbf{V}}^{H} = \sum_{i=1}^{\epsilon} \tilde{\mathbf{v}}_{i}\tilde{\mathbf{v}}_{i}^{H}$.
In (\ref{eq:eff_chb_gain_dzf_f}), given the orthonormal vectors $\{\tilde{\mathbf{v}}_{i}\}_{i=1}^{\epsilon}$, the projection of $\mathbf{h}$ onto $Sp(\tilde{\mathbf{H}})^{\perp}$ can be computed by the sum of the individual projections onto each one of the orthonormal basis \cite{Yanai2011}.
As $\|\mathbf{h}\|^2$ and $\cos^{2} \theta_{\mathbf{h} \tilde{\mathbf{v}}_{i} }$ are independent variables \cite{Au-Yeung2007}, the expected value of the effective channel gain is

\begin{equation} \label{eq:expected_value_ecgain}
    \mathbb{E}\left[  |\mathbf{h}^{H}\mathbf{w} |^2  \right] = \mathbb{E} \left[ \|\mathbf{h}\|^2 \right] \mathbb{E} \left[ \sum_{i=1}^{\epsilon} \cos^{2} \theta_{\mathbf{h} \tilde{\mathbf{v}}_{i} } \right].
\end{equation}

Given $\mathbf{h}, \tilde{\mathbf{v}}_{i} \in \mathbb{C}^{N_t \times 1}$ define the random variable $\upsilon_{i}$ as

\begin{equation} \label{eq:rv_eta_definition}
    \upsilon_{i} = \cos^{2} \theta_{\mathbf{h} \tilde{\mathbf{v}}_{i}}.
\end{equation}

According to \cite{Au-Yeung2007} the cumulative probability function of $\upsilon_{i}$ is given by $F_{\upsilon_{i}}(\upsilon_{i}) = 1 -(1-\upsilon_{i})^{N_t-1}$ and the expected value of the random variable is $\mathbb{E}\left[ \upsilon_{i} \right] = \int_{0}^{1} \upsilon_{i} f_{\upsilon_{i}}(\upsilon_{i}) d\upsilon_{i} = \frac{1}{N_t}$.

\section{}
\label{appendix:2}

\ \ \textit{Proof of the Proposition~\ref{thm:2}:}

For the sake of notation, consider the channel of the intended user $\mathbf{h}$, its aggregate interference matrix $\tilde{\mathbf{H}}$, $\mathbf{w}$ defined in (\ref{eq:define_dvsinr_precoder}), its associated matrix $\mathbf{D}$, and $\rho=P/\sigma_{n}^{2}$.
%
%
Let $\tilde{\mathbf{H}}=\mathbf{U}_{\tilde{\mathbf{H}}} \boldsymbol\Sigma_{\tilde{\mathbf{H}}} \mathbf{V}_{\tilde{\mathbf{H}}}^{H}$ be the SVD of the aggregate interference matrix, and the unitary matrix $\mathbf{U}_{\tilde{\mathbf{H}}}$ is formed by the vectors $\{\mathbf{u}_{i}\}_{i=1}^{N_t}$. The matrix $\mathbf{D}$ can be decomposed as:
\begin{equation} \label{eq:matrix_D_decomposition}
     \mathbf{D} = \sum_{i=1}^{N_t - \epsilon} \lambda_{i}(\mathbf{D}) \mathbf{u}_{i}\mathbf{u}_{i}^{H} +  \sum_{j = N_t - \epsilon +1}^{N_t} \rho \mathbf{u}_{j}\mathbf{u}_{j}^{H} = \mathbf{D}_{P} + \mathbf{D}_{Q}
\end{equation}
The effective channel gain can be expressed as

\begin{equation} \label{eq:effective_channel_gain_dvsinr}
    |\mathbf{h}^{H}\mathbf{w} |^2
    = \frac{|\mathbf{h}^{H}\mathbf{D}\mathbf{h} |^2}{ \|\mathbf{D}\mathbf{h}\|^2} = \frac{|\mathbf{h}^{H}\mathbf{D}_{P}\mathbf{h} + \mathbf{h}^{H}\mathbf{D}_{Q}\mathbf{h}|^2}{ \|\mathbf{D}_{P}\mathbf{h}\|^2 + \|\mathbf{D}_{Q}\mathbf{h}\|^2} = \| \mathbf{h} \|^2 \beta_{\mathbf{h}}
\end{equation}
where
\begin{equation} \label{eq:beta_of_h}
    \beta_{\mathbf{h}} =  \frac{\left( \sum_{i=1}^{N_t - \epsilon} \lambda_{i}(\mathbf{D}) \cos^{2} \theta_{\mathbf{h}\mathbf{u}_{i}} +\rho \sum_{j = N_t - \epsilon +1}^{N_t}  \cos^{2} \theta_{\mathbf{h}\mathbf{u}_{j}} \right)^2}{\sum_{i=1}^{N_t - \epsilon} \lambda_{i}^{2}(\mathbf{D}) \cos^{2} \theta_{\mathbf{h}\mathbf{u}_{i}} +\rho^{2} \sum_{j = N_t - \epsilon +1}^{N_t}  \cos^{2} \theta_{\mathbf{h}\mathbf{u}_{j}}}.
\end{equation}

Observe that all the eigenvalues in (\ref{eq:beta_of_h}) are multiplied by squared correlation coefficients.
Bearing in mind that $\lambda_{i}(\mathbf{D})$ and $\cos^{2} \theta_{\mathbf{h}\mathbf{u}_{i}}$ are independent, then approximating the value of $\cos^{2} \theta_{\mathbf{h}\mathbf{u}_{i}}$ by its mean (which is accurate for moderately high values of $N_t$) we get:
\begin{IEEEeqnarray}{rCl}
    \mathbb{E}   \left[ \| \mathbf{h} \|^2 \beta_{\mathbf{h}} \right]  & \approx &  \mathbb{E} \left[ \|\mathbf{h}\|^2\frac{  \left( \frac{1}{N_t} \sum_{i=1}^{N_t} \lambda_{i}(\mathbf{D})\right)^2 }{ \frac{1}{N_t} \sum_{i=1}^{N_t}\lambda_{i}^{2}(\mathbf{D}) } \right] \IEEEyessubnumber \label{eq:capacity_dvsinr_equalities2_a} \\
    & = & \mathbb{E} \left[ \|\mathbf{h}\|^2 J(\mathbf{eig}(\mathbf{D})) \right] \IEEEyessubnumber \label{eq:capacity_dvsinr_equalities2_b}
\end{IEEEeqnarray}
where (\ref{eq:capacity_dvsinr_equalities2_a}) considers the expected value $\mathbb{E}\left[ \cos^{2} \theta_{\mathbf{h}\mathbf{u}_{i}} \right] = \frac{1}{N_t}$ $\forall i$ and (\ref{eq:capacity_dvsinr_equalities2_b}) follows the definition of the Jain's index in (\ref{eq:jains_index}).

\begin{figure}[!t]
  \centering
  \includegraphics[width=3.4in]{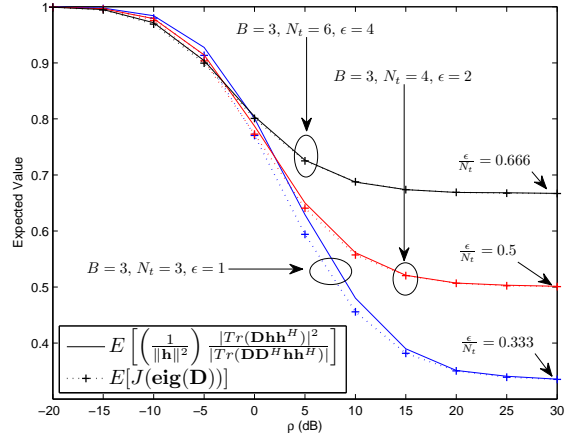}\\
  \caption{Normalized values of the effective channel gain of DVSINR precoder and its bound for $B=3$ and $N_t \in \{3, 4, 6\}$.}
  \label{fig:jain_indices}
\end{figure}

A numerical example of $\mathbb{E} \left[ |\mathbf{h}^{H}\mathbf{w} |^2 / \|\mathbf{h}\|^2 \right]$ and its approximation $\mathbb{E} \left[  J(\mathbf{eig}(\mathbf{D})) \right]$ are presented in Fig.~\ref{fig:jain_indices} for $B=3$ and $N_t \in \{3, 4, 6\}$. The curves are normalized regarding to $\|\mathbf{h}\|^2$ in order to exclusively illustrate the relation between the eigenvalues of $\mathbf{D}$. Notice that the approximation becomes tight as $\rho \rightarrow \infty$ or $\rho \rightarrow 0$, which can be realized by substituting extreme values of $\rho$ in (\ref{eq:beta_of_h}) and $J(\mathbf{eig}(\mathbf{D}))$.

\section{}
\label{appendix:3}

\ \ \textit{Proof of the Proposition~\ref{thm:3}:} Let $\tilde{\mathbf{H}}=\mathbf{U}_{\tilde{\mathbf{H}}} \boldsymbol\Sigma_{\tilde{\mathbf{H}}} \mathbf{V}_{\tilde{\mathbf{H}}}^{H}$ be the singular value decomposition (SVD) of the aggregate interference matrix of the served user. And let $\hat{\mathbf{H}} = \tilde{\mathbf{H}}\tilde{\mathbf{H}}^{H}=\mathbf{U}_{\hat{\mathbf{H}}} \boldsymbol\Sigma_{\hat{\mathbf{H}}} \mathbf{V}_{\hat{\mathbf{H}}}^{H}$ be the SVD of the Hermitian matrix $\hat{\mathbf{H}}$.
The diagonal matrix that contains the eigenvalues of $\hat{\mathbf{H}}$ can be defined from the eigenvalues of the aggregate interference matrix as
\begin{equation} \label{eq:h_hat_matrix_sigma_definition}
    \boldsymbol\Sigma_{\hat{\mathbf{H}}} = \boldsymbol\Sigma_{\tilde{\mathbf{H}}} \boldsymbol\Sigma_{\tilde{\mathbf{H}}}^{H},
\end{equation}
and the eigenvalues of the matrix $\mathbf{D}$ are given by:
\begin{equation}\label{eq:d_matrix_eigenvalues}
\lambda_{i}(\mathbf{D}) = (\rho^{-1} + [\boldsymbol\Sigma_{\hat{\mathbf{H}}}]_{ii})^{-1}.
\end{equation}
Due to the fact that $\lambda_{\min}(\hat{\mathbf{H}})$ is equal to zero with multiplicity $\epsilon$, $\lambda_{\max}(\mathbf{D}) = \rho$ with multiplicity $\epsilon$. This means that $N_t - \epsilon$ eigenvalues of $\mathbf{D}$ are bounded as $\rho \rightarrow \infty$ and $\epsilon$ are not. The Jain's index of $\mathbf{eig}(\mathbf{D})$ is such that:

\begin{IEEEeqnarray}{rCl}
    & & \lim_{\rho \rightarrow \infty} J(\mathbf{eig}(\mathbf{D})) = \lim_{\rho \rightarrow \infty} \frac{\left( \sum_{i=1}^{N_t} \lambda_{i}(\mathbf{D}) \right)^2}{ N_t\sum_{i=1}^{N_t} \lambda_{i}^2(\mathbf{D}) } \IEEEyessubnumber \label{eq:limit_jain_index_rho_grows_a} \\
    & = & \lim_{\rho \rightarrow \infty} \frac{\left( \sum_{i=1}^{N_t - \epsilon} (\rho^{-1} + [\boldsymbol\Sigma_{\hat{\mathbf{H}}}]_{ii})^{-1} +  \sum_{j = N_t - \epsilon +1}^{N_t} \rho  \right)^2}{ N_t\left( \sum_{i=1}^{N_t - \epsilon} (\rho^{-1} + [\boldsymbol\Sigma_{\hat{\mathbf{H}}}]_{ii})^{-2} +  \sum_{j = N_t - \epsilon +1}^{N_t} \rho^2 \right) } \IEEEyessubnumber \label{eq:limit_jain_index_rho_grows_b} \IEEEeqnarraynumspace \\
    & = & \lim_{\rho \rightarrow \infty} \frac{\left( \sum_{i=1}^{N_t - \epsilon} (\rho^{-1} + [\boldsymbol\Sigma_{\hat{\mathbf{H}}}]_{ii})^{-1} +  \epsilon \rho  \right)^2}{ N_t\left( \sum_{i=1}^{N_t - \epsilon} (\rho^{-1} + [\boldsymbol\Sigma_{\hat{\mathbf{H}}}]_{ii})^{-2} +  \epsilon \rho^2 \right) }
     = \frac{\epsilon}{N_t} \IEEEyessubnumber \label{eq:limit_jain_index_rho_grows_c}
\end{IEEEeqnarray}
which is illustrated in Fig.~\ref{fig:jain_indices}.

\ \ \textit{Proof of Proposition~\ref{prop:noise_interference_ratio_for_dvsinr}:} In order to simplify the notation let $\mathbf{h}_{1}$ be the channel of the user served in the local BS, with its associated matrices $\tilde{\mathbf{H}}_{1}$ and $\mathbf{D}_{1} = \mathbf{D}_{1P} + \mathbf{D}_{1Q}$ as in (\ref{eq:matrix_D_decomposition}). And let $\mathbf{h}_{2} \in \tilde{\mathbf{H}}_{1}$ be a channel vector used to create the precoding vector $\mathbf{w}_{1}$. The interference term $|\mathbf{h}_{2}^{H}\mathbf{w}_{1} |^2$ in (\ref{eq:instantaneous_sinr}) for DVSINR can be unfolded as follows:

\begin{IEEEeqnarray}{rCl}
    |\mathbf{h}_{2}^{H}\mathbf{w}_{1} |^2 & = &\frac{ |\mathbf{h}_{2}^{H}\mathbf{D}_{1}\mathbf{h}_{1}|^{2} } { |Tr(\mathbf{D}_{1}\mathbf{D}_{1}^{H}\mathbf{h}_{1}\mathbf{h}_{1}^{H})| } \IEEEyessubnumber \label{eq:interference_component_dvsinr_a} \\
    & = & \frac{ \left| \sum_{i=1}^{N_t-\epsilon} \lambda_{i}(\mathbf{D}_{1}) \mathbf{h}_{2}^{H}\mathbf{u}_{i}\mathbf{u}_{i}^{H}\mathbf{h}_{1}  \right|^2 }{ \|\mathbf{h}_{1}\|^2 \sum_{i=1}^{N_t} \lambda_{i}^{2}(\mathbf{D}_{1}) \cos^{2}\theta_{\mathbf{h}_{1}\mathbf{u}_{i}} } \IEEEyessubnumber \label{eq:interference_component_dvsinr_b} \\
    & = & \frac{ \left| \sum_{i=1}^{N_t-\epsilon} \lambda_{i}(\mathbf{D}_{1}) \left<\mathbf{h}_{2}, \mathbf{u}_{i}\right>\left<\mathbf{u}_{i},\mathbf{h}_{1}\right>  \right|^2 }{ \|\mathbf{h}_{1}\|^2 \sum_{i=1}^{N_t} \lambda_{i}^{2}(\mathbf{D}_{1}) \cos^{2}\theta_{\mathbf{h}_{1}\mathbf{u}_{i}} } \IEEEyessubnumber \label{eq:interference_component_dvsinr_c} \\
    & \leq & \|\mathbf{h}_{2}\|^2 \frac{  \left( \sum_{i=1}^{N_t-\epsilon} \lambda_{i}(\mathbf{D}_{1}) \cos\theta_{\mathbf{h}_{2}\mathbf{u}_{i}} \cos\theta_{\mathbf{h}_{1}\mathbf{u}_{i}} \right)^2 }{ \sum_{i=1}^{N_t} \lambda_{i}^{2}(\mathbf{D}_{1}) \cos^{2}\theta_{\mathbf{h}_{1}\mathbf{u}_{i}} } \IEEEyessubnumber \label{eq:interference_component_dvsinr_d} \IEEEeqnarraynumspace
\end{IEEEeqnarray}
where the numerator in (\ref{eq:interference_component_dvsinr_b}) only takes into account the basis and eigenvalues of $\mathbf{D}_{1P}$ in (\ref{eq:matrix_D_decomposition}) since $\mathbf{D}_{1Q}$ contains the basis of the null space of $\mathbf{h}_{2}$.
The result in (\ref{eq:interference_component_dvsinr_d}) obeys the triangle inequality \cite{Gentle2007} since the terms $\left<\mathbf{h}_{2}, \mathbf{u}_{i}\right>\left<\mathbf{u}_{i},\mathbf{h}_{1}\right>$ in (\ref{eq:interference_component_dvsinr_c}) are complex numbers, and by taking their associated norms and coefficients of correlation their absolute values are already computed, cf. (\ref{eq:coefficient_correlation}).

In order to define an upper bound of $ \mathbb{E} \left[ |\mathbf{h}_{2}^{H}\mathbf{w}_{1} |^2 \right]$ notice that the eigenvalues of $\mathbf{D}_{1}$ in the denominator of (\ref{eq:interference_component_dvsinr_d}) are affected by an independent random variables of the form (\ref{eq:rv_eta_definition}) with expected value $\frac{1}{ N_{t}}$.
We take the upper bound of the numerator of (\ref{eq:interference_component_dvsinr_d}) as follows. The terms $\cos\theta_{\mathbf{h}_{1}\mathbf{u}_{i}}$ and $\cos\theta_{\mathbf{h}_{2}\mathbf{u}_{i}}$ are independent so that $\mathbb{E} \left[ ( \cos\theta_{\mathbf{h}_{1}\mathbf{u}_{i}} \cos\theta_{\mathbf{h}_{2}\mathbf{u}_{i}} )^2 \right] = \mathbb{E} \left[ \cos^2\theta_{\mathbf{h}_{1}\mathbf{u}_{i}} \right] \mathbb{E} \left[ \cos^2\theta_{\mathbf{h}_{2}\mathbf{u}_{i}} \right]$. Since $0 \leq \cos^2\theta_{\mathbf{h}_{2}\mathbf{u}_{i}} \leq 1$ we have that following upper bound $\mathbb{E} \left[ ( \cos\theta_{\mathbf{h}_{1}\mathbf{u}_{i}} \cos\theta_{\mathbf{h}_{2}\mathbf{u}_{i}} )^2 \right] \leq \mathbb{E} \left[ \cos^2\theta_{\mathbf{h}_{1}\mathbf{u}_{i}} \right] = \frac{1}{ N_{t}}$.
Considering that the terms $\cos\theta_{\mathbf{h}_{1}\mathbf{u}_{i}}$ $\forall i$ are independent of $\|\mathbf{h}_{2}\|^2$ and $\mathbf{eig}(\mathbf{D}_{1})$, the expected value of the leakage is upper bounded as follows:

\begin{equation} \label{eq:interference_component_dvsinr_upper_bound}
    \mathbb{E} \left[ |\mathbf{h}_{2}^{H}\mathbf{w}_{1} |^2 \right] \leq \mathbb{E} \left[ \|\mathbf{h}_{2}\|^2 \frac{ |Tr(\mathbf{D}_{1P})|^2 }{ Tr(\mathbf{D}_{1}\mathbf{D}_{1}^{H}) } \right]
\end{equation}

For the scenarios where $N_t \geq B$ the trace ratio in (\ref{eq:interference_component_dvsinr_upper_bound}) can be approximated by dividing the largest squared eigenvalue of the numerator given by $\lambda_{\max}^{2}(\mathbf{D}_{1P}) = (\rho^{-1}+\lambda_{\min}(\tilde{\mathbf{H}}_{1}^{H}\tilde{\mathbf{H}}_{1}))^{-2}$ over the largest squared eigenvalue in the denominator $\lambda_{\max}^{2}(\mathbf{D}_{1}) =\rho^2$ which has multiplicity $\epsilon$. By considering only these eigenvalues in the ratio, the contribution of the other eigenvalues is ignored and the approximated expected value of the leakage is given by:

\begin{IEEEeqnarray}{rCl}
    \mathbb{E} \left[ |\mathbf{h}_{2}^{H}\mathbf{w}_{1} |^2 \right] & \approx & \mathbb{E} \left[ \frac{ \|\mathbf{h}_{2}\|^{2} \lambda_{\max}^{2}(\mathbf{D}_{1P}) }{  \epsilon \lambda_{\max}^{2}(\mathbf{D}_{1}) } \right] \IEEEyessubnumber \label{eq:interference_approximation_dvsinr_a} \\
    & = & \mathbb{E} \left[ \frac{ \|\mathbf{h}_{2}\|^{2} }{  \epsilon (\rho\lambda_{\min}(\tilde{\mathbf{H}}_{1}^{H}\tilde{\mathbf{H}}_{1}) + 1)^{2} } \right] \IEEEyessubnumber \label{eq:interference_approximation_dvsinr_b}
\end{IEEEeqnarray}

\begin{figure}[!t]
  \centering
  \includegraphics[width=3.4in]{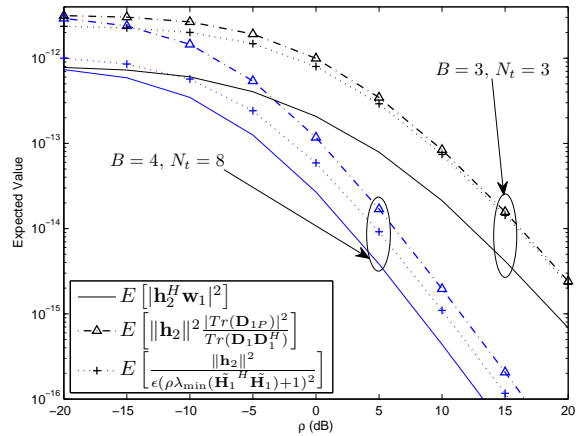}\\
  \caption{Upper bound and exact value of the average leakage $\mathbb{E} \left[ |\mathbf{h}_{2}^{H}\mathbf{w}_{1} |^2 \right]$ for $B=3$ with $N_t=3$ and $B=4$ with $N_t=8$.}
  \label{fig:leakage}
\end{figure}

A numerical example of $\mathbb{E} \left[ |\mathbf{h}_{2}^{H}\mathbf{w}_{1} |^2 \right]$, its upper bound (\ref{eq:interference_component_dvsinr_upper_bound}), and the approximation (\ref{eq:interference_approximation_dvsinr_b}) is presented in Fig.~\ref{fig:leakage}.

\section{}
\label{appendix:4}

\ \ \textit{Proof of Proposition~\ref{lemma:1}:} From (\ref{eq:matrix_D_decomposition}) it can be observed that the basis of both $Sp(\tilde{\mathbf{H}})$ and $Sp(\tilde{\mathbf{H}})^{\perp}$ are combined when forming the precoder. The value of the effective channel gain is a function of $\rho$ and for the low SNR regime $\mathbf{D}_{P}$ is dominant  while in the high SNR regime $\mathbf{D}_{Q}$ is the dominant term of $\mathbf{D}$.
The exact interaction between of the vectors $\{\mathbf{u}_{i}\}_{i=1}^{N_t}$ and $\mathbf{h}$ is given by $\beta_{\mathbf{h}}$. Observe that the components of $\mathbf{D}_{Q}$ in (\ref{eq:beta_of_h}) compute the magnitude of the projection of $\mathbf{h}$ onto each basis of $Sp(\tilde{\mathbf{H}})^{\perp}$, the exact NSP component scaled by $\rho$. The components of $\mathbf{D}_{P}$ do not represent the exact projection of $\mathbf{h}$ onto $Sp(\tilde{\mathbf{H}})$ because each one of the basis is affected by a different eigenvalue $\lambda_{i}(\mathbf{D})$. The term $|\mathbf{h}^{H}\mathbf{w} |^2$ combines a component of $\mathbf{h}$ onto $Sp(\tilde{\mathbf{H}})^{\perp}$ and weighted components of $\mathbf{h}$ onto the basis of $Sp(\tilde{\mathbf{H}})$.

The intuition behind the heuristic metric (\ref{eq:metric_of_selection}) is that the effective channel gain is bounded as follows:
\begin{equation} \label{eq:bounds_of_ecgain}
    \|\mathbf{Q}_{\mathbf{h}}\mathbf{h} \|^2 \leq  |\mathbf{h}^{H}\mathbf{w} |^2 \leq \|\mathbf{h}\|^2
\end{equation}

This means that one can always take into account the magnitude of $\|\mathbf{Q}_{\mathbf{h}}\mathbf{h} \|^2$ and the component $\|\mathbf{P}_{\mathbf{h}}\mathbf{h} \|^2$ should be modified by a monotonic decreasing function of $\rho$ with values in the range $[0,1]$.
By observing that $\beta_{\mathbf{h}}$ is the ratio of the squared combination of the eigenvalues of $\mathbf{D}$ over the combination of its squared eigenvalues, the function (\ref{eq:heuristic_alpha_dvsinr}) is defined by the quotient of $\lambda_{\min}^{2}(\mathbf{D}) = (\rho^{-1} + \lambda_{\max}(\tilde{\mathbf{H}}^{H}\tilde{\mathbf{H}}))^{-2}$ over $\lambda_{\max}^{2}(\mathbf{D}) = \rho^2$. The objective of such ratio is to measure how much the maximum and minimum eigenvalues of $\mathbf{D}$ spread out as a function of $\rho$. Observe that as $\rho \rightarrow 0$ the value of (\ref{eq:heuristic_alpha_dvsinr}) goes to 1 and when $\rho \rightarrow \infty$ the function goes to zero.

\ \ \textit{Proof of Proposition~\ref{prop:over_approximation_nsp}:}

Let $\tilde{\mathbf{V}}_{bk_{b}}(\mathcal{G}_{l})=null(\tilde{\mathbf{H}}_{bk_{b}}(\mathcal{G}_{l}))$ be the matrix that contains the orthonormal basis of the null space of $\tilde{\mathbf{H}}_{bk_{b}}(\mathcal{G}_{l})$ and $\tilde{\mathbf{v}}_{i}$ is its $i$th column vector with $i \in \{1,\ldots,\epsilon \}$. The NSP can be computed as $\|\mathbf{h}_{bk_{b}} \mathbf{Q}_{\mathbf{h}_{bk_{b}}}\|^2 = \|\mathbf{h}_{bk_{b}}\|^2 \sum_{i=1}^{\epsilon} \cos^{2}\theta_{\mathbf{h}_{bk_{b}}\tilde{\mathbf{v}}_{i}}$ (see Appendix~\ref{appendix:1}). Recall that $\|\mathbf{h}_{bk_{b}}\|^2$ and $\cos^{2}\theta_{\mathbf{h}_{bk_{b}}\tilde{\mathbf{v}}_{i}}$ are independent variables \cite{Au-Yeung2007} and the factors of the product in the RHS of (\ref{eq:upper_bound_app_nsp}) are independent. Assuming that the components of $\tilde{\mathbf{H}}_{bk_{b}}(\mathcal{G}_{l})$ are i.i.d., we have that:
    \begin{equation} \label{eq:mean_of_app_nsp}
        \mathbb{E}\left[ \prod_{i_{b}\neq k_{b}, i_{b} \in \mathcal{G}_{l}} \sin^{2}\theta_{\mathbf{h}_{bk_{b}} \mathbf{h}_{bi_{b}}} \right] = \left(1-\frac{1}{N_t}\right)^{(B-1)}
    \end{equation}
and $\frac{\epsilon}{N_t} \leq  \left(1-\frac{1}{N_t}\right)^{(B-1)}$ with equality when $B=2$ for a given $N_t$. Notice that equality is asymptotically attained for a fixed value of $B$ when $N_t \rightarrow \infty$.


\ifCLASSOPTIONcaptionsoff
  \newpage
\fi

%
%

\bibliographystyle{IEEEtran}
\bibliography{multicell_user_selection_tvt}

\includepdf[pages={-}]{double_column.pdf}

\end{document}